%% file: journalv19.tex
\documentclass[draftcls,onecolumn]{IEEEtran}
 
\usepackage{amsmath, amsfonts, amssymb}
\usepackage{bbm}
\usepackage{bbold}
\usepackage{graphicx,dblfloatfix}
\usepackage[space]{cite}
\usepackage{hyperref}
\usepackage{color}

\usepackage{pgfplotstable}
 
\newtheorem{theorem}{Theorem}
\newtheorem{corollary}[theorem]{Corollary}
\newtheorem{lemma}[theorem]{Lemma}
\newtheorem{proposition}[theorem]{Proposition}
\newtheorem{definition}{Definition}
\newtheorem{remark}{Remark}

\newcommand{\mbf}[1]{\mathbf{#1}}
\newcommand{\set}[1]{\mathcal{#1}}

\newcommand{\bfZ}{\mathbb{V}}
\newcommand{\bfV}{\mathbb{V}}
\newcommand{\V}{\bfV}

\renewcommand{\Pr}{\mathbb{P}}

\newcommand{\M}{\mathsf{M}}
\newcommand{\C}{\mathsf{C}}
\renewcommand{\d}{\mbf{d}}
\newcommand{\Pe}{{\mathsf{P}_\text{e}}}
\newcommand{\Qkone}{\mathcal{Q}_{k-1}^{\textnormal{dist}}}
\newcommand{\K}{\set{K}}

\newcommand{\A}{\textnormal{A}}
\newcommand{\B}{\textnormal{B}}

\newcommand{\capa}{\mathbf{C}}
\allowdisplaybreaks[4] 
\graphicspath{{.}{./Figures/}} 

\renewcommand{\c}{\textnormal{c}}

\begin{document}
\title{Benefits of Cache Assignment on Degraded Broadcast Channels}
\author{Shirin Saeedi Bidokhti, Mich\`{e}le Wigger, and Aylin Yener
\thanks{S.~Saeedi~Bidokhti is with the Department of Electrical Engineering at Stanford University, saeedi@stanford.edu.  S.~Saeedi~Bidokhti is supported by the Swiss National Science Foundation fellowship no. 158487. M.~Wigger is with LTCI, Telecom ParisTech, Universit\'e Paris-Saclay, 75013 Paris, michele.wigger@telecom-paristech.fr. A.~Yener is with the Department of Electrical Engineering, School of Electrical Engineering and Computer Science at The Pennsylvania State University and the Department of Electrical Engineering at Stanford University, yener@engr.psu.edu, yener@stanford.edu. Parts of the material in this paper have been submitted to  the  \emph{IEEE International Conference on Communications}, Paris, May 2017, and  to the \emph{IEEE International Symposium on Information Theory}, Aachen, Germany, June 2017.}}
\maketitle


\begin{abstract}
Degraded $K$-user broadcast channels (BC) are studied when receivers are facilitated with cache memories. Lower and upper bounds are derived on the {capacity-memory tradeoff}, i.e., on the largest rate of reliable communication over the BC as a function of the receivers' cache sizes, and the bounds are shown to match for some special cases. The lower bounds are achieved by two new coding schemes that benefit from \emph{non-uniform cache assignment}.  Lower  and upper bounds  are also established  on the global capacity-memory tradeoff, i.e., on the largest capacity-memory tradeoff that can be attained by optimizing the receivers' cache sizes subject to a {total} cache memory budget. The bounds coincide when the total cache memory budget is sufficiently small or sufficiently large, characterized in terms of the BC statistics. 
For small cache memories, it is optimal to assign all the cache memory to the weakest receiver. In this regime, the global capacity-memory tradeoff grows as the total cache memory budget divided by the number of files in the system. In other words, a \emph{perfect global caching gain} is achievable in this regime and the performance corresponds to a system where all cache contents in the  network are available to all receivers. 
For large cache memories, it is optimal to assign a positive cache memory to every receiver such that the weaker receivers are assigned larger cache memories compared to the stronger receivers. In this regime, the growth rate of the  global capacity-memory tradeoff is further divided by the number of users, which corresponds  to a local caching gain.
Numerical indicate suggest that a uniform cache-assignment of the total cache memory is suboptimal in all regimes unless the BC is completely symmetric. For erasure BCs, this claim is proved analytically in the regime of small cache-sizes.

\end{abstract}
\section{Introduction}
Storing popular contents at or close to the end users  {improves} the network performance during peak-traffic time. The main challenge is that the contents have to be cached before knowing which files the users will request in the peak-traffic period. A conventional approach is to store the same popular contents in the  cache memories of the users. This allows  the receivers to locally retrieve the contents  without burdening the network. However, further caching gains, i.e., the so called \emph{global caching gains}, are possible if different contents are stored at different users~\cite{maddahali_niesen_2014-1}. Specifically, a careful design of the cache contents creates coding opportunities to simultaneously serve multiple users  during the peak-traffic periods, henceforth called the \emph{delivery phase}.

\begin{figure}[t!]
 \begin{center}
 	\includegraphics[width=.58\textwidth]{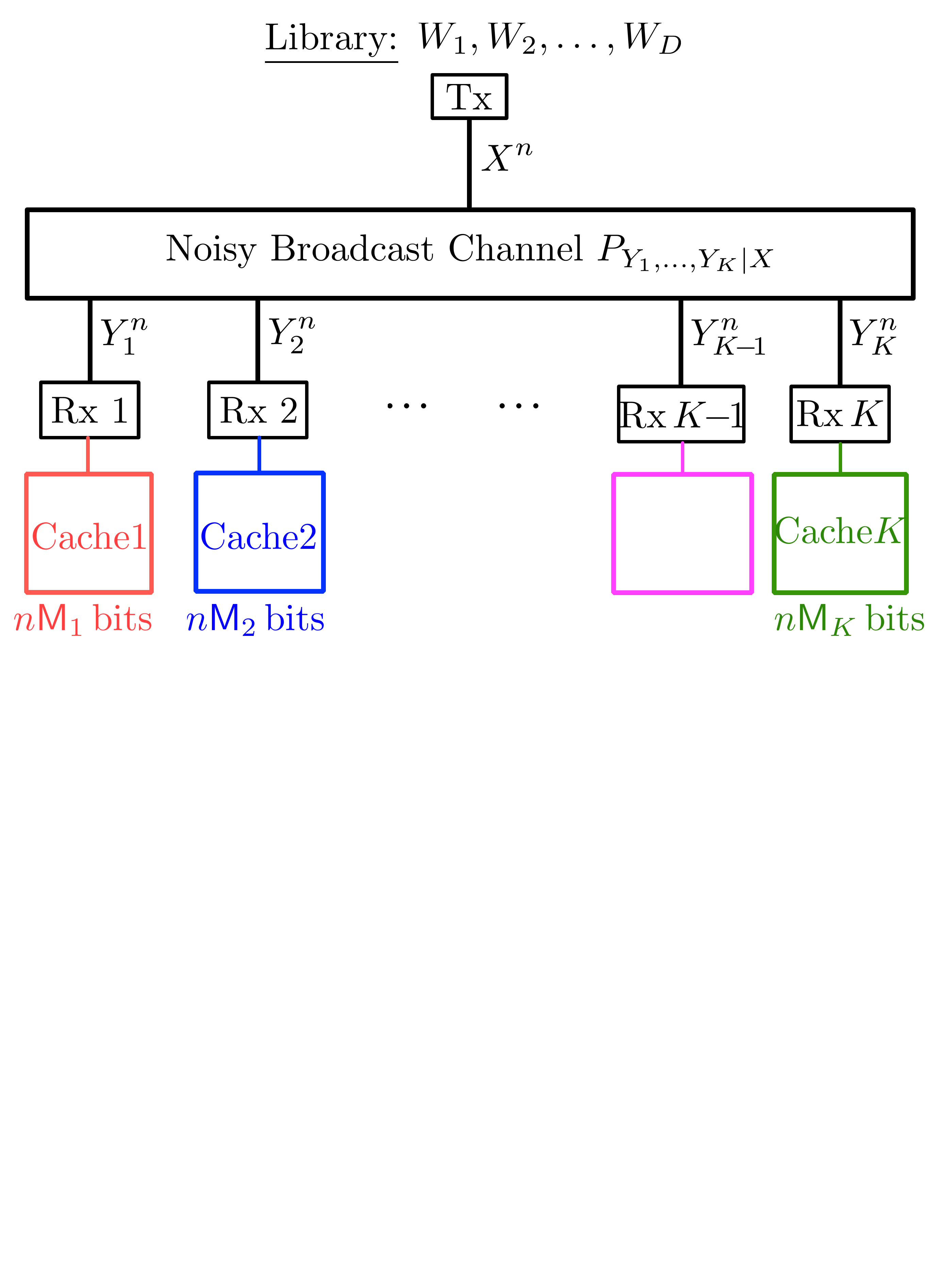}
 	\vspace{-5.6cm}
 	\caption{Noisy broadcast channel with cache memories at the receivers.}
 	\label{fig:model}
 \end{center}

 \end{figure}

In this paper, we focus on the scenario depicted in Figure~\ref{fig:model}.
A transmitter communicates with receivers $1,\ldots, K$ which are equipped with cache memories. The delivery-phase  communication takes place over a noisy broadcast channel (BC) where the receivers have access to cache memories. 
 The  BC-model has previously been studied in \cite{timowigger-2015-1,saeedi_wigger_timo-2016-ISIT, saeeditimowigger-IT, saeediwiggertimo-turbo,huang_wang_ding_yang_zhang_2015,wang_xian_liu_2015,hassanzadeherkipllorcatulino-2015,yangngokobayashi-2016,ghorbelkobayashiyang-2016,ghorbelkobayashiyang-2016-2,ghorbelkobayashiyang-2015,tulino_jccs,hachemkaramchandanidiggavi-2015-2,zhangelia-2015,zhangengelmannelia-2015,zhangelia-2016,zhangelia-2016-2,yicaire-2016}.
The simplified version where the BC is a  common noise-free bit-pipe to all users was analyzed in\cite{maddahali_niesen_2014-1,chenfanletaief-2014,tian-2015,wan_tuninetti_piantanida-2015,wan_tuninetti_piantanida-2016,sengupta_tandon_clancy-2015, amirigunduz-2016,tianchen-2016,sahraeigastpar-2016, tian-2016,YuMA:16,wanglimgastpar-2016,ghasemi_ramamoorthy,ghasemi_ramamoorthy-2016, AmiriYangGunduz,wangLiTianLiu,Timo-Mar-2016-C,Timo-Oct-2016-A, wang_lim_gastpar_2015, limwanggastpar-2016 } under the assumption that all receivers have equal cache sizes, and in  \cite{AmiriYangGunduz,wangLiTianLiu} under the assumption that   various receivers  have different cache sizes. 
Caching was studied for many other scenarios, e.g.,  for  interference networks \cite{maddahaliniesen-2015-1,naderializadehmaddahaliavestimehr-2016,pooyaabolfazlhossein-2015}, hierarchical networks \cite{nikhil1 ,hachemkaramchandanidiggavi-2014-2,hachemkaramchandanidiggavi-2015}, and cellular networks~\cite{WiggerTimoShamai16,ntranosmaddahalicaire-2015,pooyaabolfazlhossein-2015,ugurawansezgin-2015,parksimeoneshamai-2016,tandonsimeone-2016,azarisimeonespagnolinitulino-2016,pengyanzhangwang-2015}.

In \cite{timowigger-2015-1, saeeditimowigger-IT, saeedi_wigger_timo-2016-ISIT}  the gains of caching in noisy broadcast networks are investigated. Specifically, we  have proposed  a joint cache-channel coding scheme  and focused on  erasure BCs with two sets of receivers: a set of cache-aided weak receivers  (where each channel has the same erasure probability)  and a set of strong receivers without cache memories (where each channel has the same erasure probability). Previous works have adapted a separate cache-channel coding architecture where the encoders (resp. decoders) consist of a cache encoder (resp. decoder) that only exploits the cache contents and a channel encoder (resp. decoder) that only exploits the channel statistics; see Figure~\ref{fig:separate}. By contrast, in a joint cache-channel coding scheme, the encoders and decoders simultaneously exploit the knowledge of the channel statistics and the cache contents, leading to improved performance.
 {\begin{figure}[h!]
	\vspace{-2mm}
	\centering
	\qquad\qquad\qquad\qquad\qquad\includegraphics[width=1.1\columnwidth]{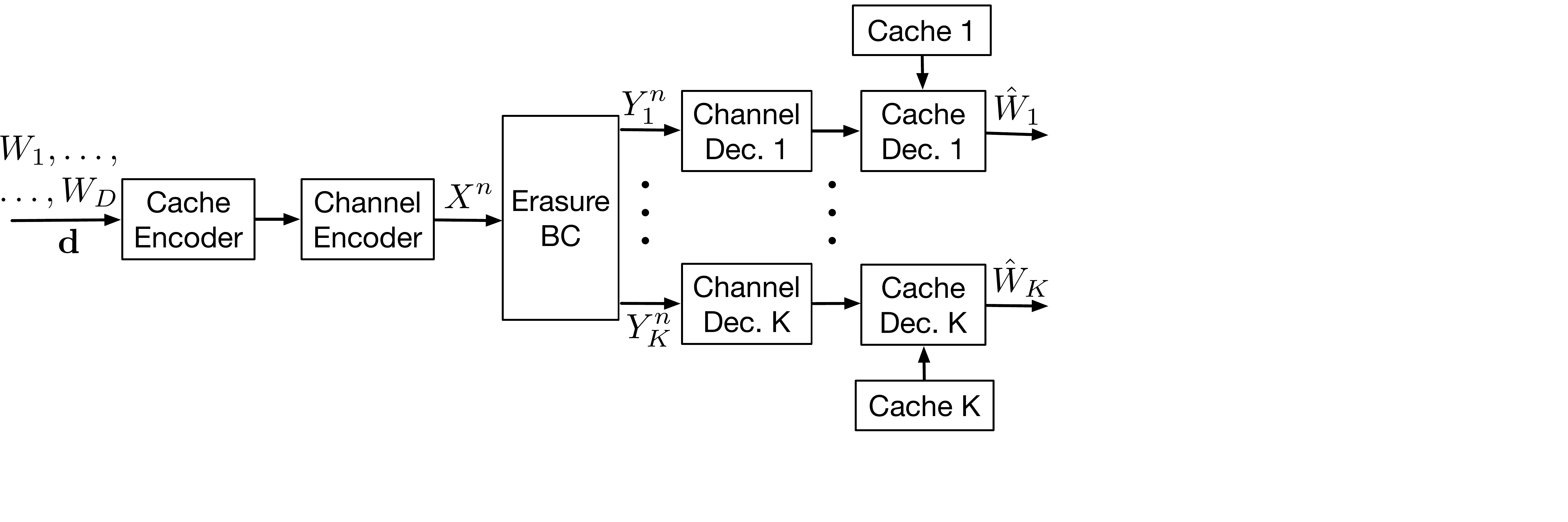}	
	\vspace{-1.8cm}
	\caption{Separate cache-channel coding architecture.}
	\label{fig:separate}
	\end{figure}}

The  joint cache-channel coding scheme in  \cite{timowigger-2015-1, saeeditimowigger-IT, saeedi_wigger_timo-2016-ISIT} loads (piggybacks) the information that is intended for the strong receivers, but is already cached at the weaker receivers,  onto the information that is communicated to the weak receivers\footnote{ {The proposed piggyback coding can be seen as a simplified version without binning etc. of   ``Slepian-Wolf coding over broadcast channels" in \cite{Tuncel-Apr-2006-A}, which applies to more general scenarios.}}. When the  {rate of the} piggybacked information is modest, this can be done without harming the decoding performance at the  {strong} receivers. In some sense, piggyback coding  provides the stronger receivers virtual access to the weaker receivers' cache-memories as if these cache contents were locally present at the stronger receivers.

 The previous works  \cite{timowigger-2015-1,saeeditimowigger-IT,saeedi_wigger_timo-2016-ISIT} have shown that when different receivers have different channel statistics, then assigning larger cache memories to the weaker receivers significantly improves the performance  {compared to} the traditional uniform cache assignment. In addition to mitigating the rate-bottleneck at the weaker receivers,  non-uniform cache assignment allows  to achieve new global caching gains by the means of \emph{joint cache-channel coding} \cite{timowigger-2015-1, saeeditimowigger-IT,saeedi_wigger_timo-2016-ISIT,tulino_jccs}.

Motivated by the new gains of caching in noisy broadcast networks, in this work, we address the problem of efficient cache assignment in  broadcast networks and devise two joint cache-channel coding schemes by using piggyback coding,  superposition coding, and  coded caching.
		\subsection{Main Contributions and Implications}
		The main contributions of the paper are as follows:
\begin{itemize}
	\item \textit{Superposition-Piggyback Coding}: We generalize the piggyback-coding scheme of \cite{saeeditimowigger-IT}, that is specific for erasure BCs, to  arbitrary BCs with a cache memory only at the weakest receiver and account for different channel qualities in the network by employing superposition coding. We show that this scheme is optimal for small cache memory sizes.
	
	\item \textit{Generalized Coded-Caching}: The coded-caching scheme in \cite{maddahali_niesen_2014-1} is generalized to \emph{noisy BCs} with unequal cache sizes. The scheme is optimal for a particular cache assignment. 
	
	\item \textit{A New Converse Result:} A general converse result is provided for degraded BCs with arbitrary cache sizes at the receivers. It strictly improves over the existing converse results for degraded BCs in \cite{saeeditimowigger-IT,saeediwiggertimo-turbo} and for the  noise-free bit-pipe model in  \cite{maddahali_niesen_2014-1, ghasemi_ramamoorthy, sengupta_tandon_clancy-2015, ji_tulino_llorca_caire_2015, wanglimgastpar-2016}.

				\item \textit{Global Capacity-Memory  Tradeoff:} Lower and upper bounds are derived on the global capacity-memory tradeoff. They are shown to match  when the total available cache memory is  small or large.  
Suboptimality of the popular approach of assigning equal cache memory to all receivers is proved analytically for erasure BCs in the small cache size regime and shown numerically for erasure and Gaussian BCs  in all regimes of cache sizes.		
\end{itemize}

More specifically, we first propose a coding scheme that we call \emph{superposition piggyback-coding} by assuming that only the weakest receiver has a cache memory. Using this scheme all receivers gain virtual access to the weakest receiver's cache memory as if the cache contents were locally present at each  of these receivers.
	
	 {The second scheme generalizes the coded caching in \cite{maddahali_niesen_2014-1} to account for different channel statistics  and different cache sizes at the receivers. 
We assign larger cache sizes to the weaker receivers and use piggyback coding  to transmit higher rates of information  to the stronger receivers without harming the communication to the weaker receivers}. As a consequence, the amount of the virtual cache memory that is provided to the stronger receivers increases compared to the original coded-caching scheme, resulting in an improved performance.    
	
The performance criterion of interest in this paper is the \emph{capacity-memory tradeoff}. That is, the largest  rate,  as a function of the available cache memories, so that  the transmitter can reliably  send the messages demanded by the receivers over the noisy BC.

	We  present a new upper bound on the capacity-memory tradeoff of degraded BCs\footnote{Since for our purposes only the conditional marginal distributions matter, it suffices that the BC is \emph{stochastically degraded} \cite{book:gamal}.)} that improves  the previous upper bound in \cite{saeediwiggertimo-turbo, saeeditimowigger-IT}.  {Using the upper bound, we show the optimality of the superposition piggyback-coding scheme when only the weakest receiver has a cache memory and its size is below a certain threshold that depends on the BC statistics}.  {Moreover, we show that the generalized coded-caching scheme is optimal for a particular cache assignment}. 

	 {When the BC is a  noise-free bit-pipe, the upper bound on the capacity-memory tradeoff leads to a  lower bound on the delivery rate-memory tradeoff that improves the previous lower bounds in \cite{maddahali_niesen_2014-1, ghasemi_ramamoorthy, sengupta_tandon_clancy-2015, ji_tulino_llorca_caire_2015, wanglimgastpar-2016}. }
 
 {The  upper bound is asymmetric in the cache sizes: the cache memories at the weaker receivers increase the upper bound more  than the cache memories at the stronger receivers. }
In this sense, the upper bound reinforces the intuition obtained from the lower bounds that  the  capacity-memory tradeoff increases when larger cache memories are assigned to the weaker receivers as compared to the stronger receivers. To make this statement more precise, we derive upper and lower bounds on the  \emph{global capacity-memory tradeoff}, where one is allowed to optimize over the cache assignment subject to a global cache constraint. The lower bound is obtained using the following cache-assignment strategy and coding schemes: 
\begin{itemize}
	\item  For a small total cache-size $\M$, all of it is assigned to the weakest receiver, and superposition piggyback-coding is applied. This strategy is optimal in the small total cache-size  regime and achieves a global capacity-memory tradeoff that grows as $ \frac{\M}{D}$, where $D$ denotes the total number of files. Thus, in this regime a \emph{perfect global caching gain} is achieved,  i.e.,  the same performance as in a systems where \emph{all} cache memories in the network are accessible by \emph{all} the receivers.
		\item 
	For moderate total cache-size $\M$, generalized coded-caching with parameters $t=1,\ldots, K-1$ and the corresponding cache-assignments are employed. 	The larger the total cache-size, the larger the parameter $t$ needs to be chosen. However, the larger $t$,  the smaller the global caching gain, since  with increasing $t$ the overlap of the different cache contents increases as well, and duplicated cache contents cannot provide global caching gain. 	
 
		\item When the total cache-size $\M$  equals the total cache memory of generalized coded-caching with parameter $t=K-1$, then generalized coded-caching is optimal. 		
		 For total cache memories  exceeding this threshold, it is optimal to uniformly assign the \emph{additional} cache memory across the $K$ receivers. This additional cache memory can only bring  \emph{local caching gain} and the same content can be stored at all the $K$ receivers. In other words, for total cache memory exceeding a threshold, the global capacity-memory tradeoff grows as $\frac{1}{K} \cdot \frac{\M}{D}$.

\end{itemize}
Finally, this paper proves analytically that for erasure BCs a uniform cache allocation is \emph{strictly suboptimal} in the regime of small cache memories, unless all receivers have same channel statistics. Numerical simulations show that the same holds for all regimes of cache memory and also for Gaussian BCs. 

\subsection{Notation} 
\label{sec:notation}
Random variables are denoted by uppercase letters, e.g. $A$, their alphabets by matching calligraphic font, e.g. $\set{A}$, and elements of an alphabet by lowercase letters, e.g. $a \in \set{A}$. 
We also use uppercase letters for deterministic quantities like rate $R$, capacity $\C$, number of users $K$, cache size $\M$, and number of files in the library $D$. 
Vectors are identified by bold font symbols, e.g., $\mathbf{a}$, and matrices by the font $\mathsf{A}$. 
We use the shorthand notation $A^n$ for  $(A_1,\ldots, A_n)$. 
The Cartesian product of $\set{A}$ and $\set{A}'$ is $\set{A} \times \set{A}'$, and the $n$-fold Cartesian product of $\set{A}$ is $\set{A}^n$. 
$|\set{A}|$ denotes the cardinality of $\set{A}$.

Finally, for indices $w_1$ and $w_2$ taking value in $\big\{1,\ldots, \lfloor 2^{\ell_1}\rfloor\big\}$ and $\{1,\ldots, \lfloor 2^{\ell_2}\rfloor\}$, respectively, we denote by 
\begin{equation*}
w_1\bigoplus w_2
\end{equation*} the index in $\{1,\ldots, \lfloor 2^{\ell_{\max}}\rfloor\}$  that corresponds to the XOR of  the length-$\ell_{\max}$ binary representations of  $w_1$ and $w_2$, where $\ell_{\max}:=\max\{\ell_1, \ell_2\}$. 

We will be using the abbreviation {i.i.d.} for \emph{independent and identically distributed}.

\subsection{Outline} 
The remainder of the paper is organized as follows. Section~\ref{sec:model} describes the problem setup. Section~\ref{sec:preliminaries} recalls known results for the scenario without  cache memories. The main results of this paper are described in Sections~\ref{sec:lower} and \ref{sec:upper}, followed by applications of these results to erasure and Gaussian BCs, see Section~\ref{sec:examples}. The paper is concluded with a summary and conclusions, Section~\ref{sec:summary} and various technical appendices contain  the proofs of the results in Sections~\ref{sec:upper} and \ref{sec:examples}. 

\section{Problem Definition}\label{sec:model}

Consider a transmitter and receivers~$1,\ldots,K$. The transmitter has access to a library with 
$D$ independent messages, $W_1,\ldots, W_D$, each distributed uniformly  over the set
$
\big\{1,\ldots, \lfloor 2^{n R} \rfloor \big\}.
$
Here,  $R \geq 0$ denotes the rate of transmission and $n$ is the transmission blocklength. We assume that there are more messages than receivers: 
\begin{equation}
D \geq K.
\end{equation}

Each receiver $k \in \K:=\{1,\ldots,K\}$  is equipped with a cache of size $\M_k\geq 0$. 
Communication takes place in two phases. For the first,  i.e., the placement phase, the transmitter chooses caching functions 
\begin{IEEEeqnarray}{rCl}
	g_k \colon   \{1,\ldots, \lfloor 2^{nR} \rfloor \}^D
	\to \big\{1,\ldots, \lfloor 2^{n \M_k}\rfloor \big\}, \qquad k\in\K,
\end{IEEEeqnarray} and places 
\begin{equation}\label{eq:caching}
\bfV_k := g_k(W_1, \ldots, W_D)
\end{equation}
in receiver~$k$'s cache. This phase takes place in a noiseless fashion.\footnote{Following previous works on caching systems, we will also assume that the placement phase takes place in low-traffic hours with abundance of bandwidth resources, and can be considered noiseless.}

The subsequent delivery phase takes place over a  \emph{degraded}  BC \cite{gallager74} with finite input alphabet $\mathcal{X}$, finite output alphabets $\mathcal{Y}_1, \ldots, \mathcal{Y}_K$,\footnote{The results of this paper readily extends to continuous alphabets. We will consider Gaussian BCs in Section~\ref{sec:Gaussian}.} and 
channel transition law 
\begin{equation}
\Gamma(y_1,\ldots, y_K|x), \textnormal{ for }x \in\set{X}, y_1 \in \set{Y}_1, \ldots, y_K\in\set{Y}_K
\end{equation}
which decomposes as 
\begin{equation}\label{eq:channel}
\Gamma(y_1,\ldots, y_K|x)=\Gamma_K(y_K|x) \cdot \Gamma_{K-1}(y_{K_1}|y_K) \cdots  \Gamma_{1}(y_{1}|y_2) .
\end{equation}
Without loss in generality, we order the receivers from the weakest to the strongest.

At the beginning of the delivery phase, each receiver~$k$ demands message $W_{d_k}$, $d_k\in\set{D}:=\{1,\ldots,D\}$. Transmitter and all the 
receivers are informed of the demand vector 
\[
\d:=(d_1, \ldots, d_K).
\] Using this information, the transmitter forms the channel input sequence $X^n=(X_1,\ldots, X_n)$ as
\begin{equation}\label{eq:encoding}
X^n = f_\d(W_1,\ldots,W_D)
\end{equation}
for some encoding function $f_\d :  
\{1,\ldots,\lfloor 2^{nR} \rfloor \}^D
\to 
\set{X}^n.$

Receiver $k\in\K$ observes the channel output sequence $Y_k^n := (Y_{k,1},$ $\ldots,Y_{k,n})$.  Given the demand vector~$\d$, 
cache content~$\bfZ_k$, and 
channel outputs $Y_k^n$, it produces its estimate of the desired  message $W_{d_k}$,
\begin{equation}\label{eq:decoding}
\hat{W}_k := \varphi_{k,\d}(Y_k^n, \bfZ_k),
\end{equation}
by means of a decoding function
\begin{IEEEeqnarray}{rCl}
	\varphi_{k,\d} \colon \set{Y}_k^n \times \big\{1, \ldots, \lfloor 2^{n\M_k}\rfloor \big\} \to \{1,\ldots,\lfloor 2^{nR}\rfloor \}.
\end{IEEEeqnarray}

The worst-case probability of error at any receiver and  any demand $\d$  is given by
\begin{equation}\label{eq:Pe}
\Pe := 
\Pr\bigg[\ \bigcup_{\d \in \set{D}^K} 
\bigcup_{k = 1}^K
\big\{ \hat{W}_k \neq W_{d_k} \big\}\
\bigg].
\end{equation}

A rate-memory tuple $(R,\M_1,\ldots, \M_K)$  is  \emph{achievable} if for any $\epsilon >  0$ there exists a sufficiently large blocklength~$n$ and caching, encoding, and decoding functions as in \eqref{eq:caching}, \eqref{eq:encoding}, and \eqref{eq:decoding} so that $\Pe \leq \epsilon$.
\begin{definition}
	The \emph{capacity-memory tradeoff $\C(\M_1,\ldots,\M_K)$} is the largest rate $R$ for which the  rate-memory tuple  $(R,\M_1,\ldots, \M_K)$ is achievable:
	\begin{equation*}
	\C(\M_1,\ldots,\M_K):= \sup \{ R \colon (R,\M_1,\ldots, \M_K) \textnormal{ achievable}\}.
	\end{equation*}
\end{definition}

Our main goal in this paper is to optimize the cache assignment $(\M_1,\ldots, \M_K)$  to attain the largest capacity-memory tradeoff $\C(\M_1,\ldots,\M_K)$ under the total cache  constraint: 
\begin{equation}\label{eq:total_cache}
\sum_{k=1}^K \M_k \leq \M.
\end{equation}


\begin{definition}
	The \emph{global capacity-memory tradeoff $\C^\star(\M)$} is defined as:
	\begin{equation}\label{eq:global}
	\C^\star(\M):= \max_{\substack{\M_1,\ldots, \M_K>0\colon \\ \sum_{k=1}^K \M_k \leq \M}} \C(\M_1,\ldots, \M_K). 
	\end{equation}
\end{definition}


\begin{remark}
	The global capacity memory tradeoff  depends on the BC law $\Gamma(y_1,\ldots, y_K|x)$ only through its marginal conditional laws. All our results  thus also apply to 
	\emph{stochastically degraded BCs}.
\end{remark}

\subsection{Minimum Delivery Rate}
Previous works on caching that modelled the BC as a  noise-free bit-pipe, e.g., \cite{maddahali_niesen_2014-1}, adopted a ``source-coding perspective"  {as opposed to a ``channel coding perspective" as we have presented above. In the source coding perspective,}  each message is an $F>0$ bits packet, the delivery communication consists of  $\rho \cdot F$ channel uses, and  receiver~$k$ has $m_k F$ bits of cache memory, $k=1,\ldots,K$. 
The \emph{delivery rate} $\rho$ is said to be achievable given normalized memory sizes $m_1,\ldots, m_K$ if there exist caching, encoding, and decoding functions such that   the probability of error in \eqref{eq:Pe} tends to 0 as $F \to \infty$. 

The following correspondence holds between the two perspectives:   
\begin{IEEEeqnarray*}{Cl}
 R \textnormal{ achievable with } (\M_1, \ldots, \M_K) \\
 \textnormal{under  the ``channel-coding perspective"}\\[1ex]
\Longleftrightarrow \\[1.2ex]
\rho=\frac{1}{R}\  \textnormal{ achievable with } \bigg(m_1 =\frac{\M_1}{R}, \ldots, m_K= \frac{\M_K}{R}\bigg)\\
 \textnormal{under  the ``source-coding perspective"}.
\end{IEEEeqnarray*}

For simplicity, we will adopt the ``source-coding perspective" in Section~\ref{sec:sourcecoding} where we specialize the new upper bound on the capacity-memory tradeoff $\C(\M_1, \ldots, \M_K)$ to the  noise-free bit-pipe model with uniform cache assignment in \cite{maddahali_niesen_2014-1}. For other BCs, we use the ``channel-coding perspective"  {in line with similar setups in network information theory}.

\section{Preliminaries: Capacities without Cache Memories}\label{sec:preliminaries}
In the absence of cache memories, 
$$\M_1=\ldots= \M_2=0,$$ the capacity-memory tradeoff $\C(\M_1=0, \ldots, \M_K=0)$ is well known: It is the largest symmetric rate $R$ with which $K$ independent messages can be reliably sent to the $K$ receivers. 
I.e., 
\begin{equation}
\C(\M_1=0, \ldots, \M_K=0)= \C_\K \label{eq:C0op}
\end{equation}
where  \cite{gallager74}:
	\begin{IEEEeqnarray}{rCl}\label{eq:common_msg}
		\C_\K&:=&  \max  \; \min\big\{ I(U_1; Y_1) ,\ I(U_2;Y_2|U_1), \ \ldots,\ I(U_{K-1};Y_{K-1}|U_{K-2}) ,\ I(X; Y_K|U_{K-1})\big\},\IEEEeqnarraynumspace
	\end{IEEEeqnarray}
and the maximization in \eqref{eq:common_msg} is over all  random tuples $U_1, \ldots, U_{K-1},X, Y_1, \ldots, Y_K$ forming the Markov chain
\begin{subequations}\label{eq:auxiliaries}
	\begin{equation}\label{eq:markovK}
	U_1 - U_2 - \cdots -  U_{K-1} - X - (Y_1, \ldots, Y_K)
	\end{equation}
	satisfying the channel transition law
	\begin{equation}\label{eq:markov_channel}
	P_{Y_1\ldots Y_K|X}(y_1,\ldots, y_K|x)= \Gamma(y_1,\ldots, y_K|x).
	\end{equation}
\end{subequations}

To present the results in this paper, we will need the \emph{capacity region} without cache memories of the BC to a subset of the receivers
\begin{align}
\set{S}:=\{j_1, \ldots, j_{|\set{S}|}\} \subseteq \K,\quad  {\ j_1<\dots < j_{|\set{S}|}}.\label{Aylin}
\end{align}
This capacity region $\capa_{\set{S}}$ \cite{gallager74} is given by 
the set of all nonnegative rate-tuples $(R_1, \ldots, R_{|\set{S}|})$ for which there exist  random variables $U_1, \ldots, U_{|\set{S}|-1},X, Y_{j_1}, \ldots, Y_{j_|\set{S}|}$ satisfying \eqref{eq:markov_channel} and forming the Markov chain
\begin{equation}\label{eq:markovS}
U_{1} - U_{2} - \cdots -  U_{|\set{S}|-1} -X - \big(Y_{j_1}, \ldots, Y_{j_{|\set{S}|}}\big),
\end{equation}
such that the following  conditions hold: 
\begin{subequations}\label{eq:capacity_region}
	\begin{IEEEeqnarray}{rCl}
		R_1 & \leq & I(U_{1}; Y_{j_1}), \\
		R_k & \leq & I(U_k;Y_{j_k}|U_k), \quad k\in \{2,\ldots, |\set{S}|-1\}, \\
		R_{|\set{S}|} & \leq & \ I\big({X}; Y_{j_{|\set{S}|}}\big|U_{|\set{S}|-1}\big). 
	\end{IEEEeqnarray} 
\end{subequations}
We  denote by $\C_{\set{S}}$ the largest symmetric rate $R\geq 0$ in $\capa_\set{S}$: 
\begin{equation}
\C_{\set{S}} := \max_{R\geq 0} \{R \colon (R, \ldots, R)\in\capa_{\set{S}}\}.
\end{equation}
It equals 
\begin{IEEEeqnarray}{rCl}
\C_\set{S}&=&  \max  \; \min\big\{ I(U_1; Y_{j_1}) ,\ I(U_2;Y_{j_2}|U_1), \ \ldots,\  I(U_{|\set{S}|-1};Y_{j_{|\set{S}|-1}}|U_{|\set{S}|-2}) ,\ I(X; Y_{j_{|\set{S}|}}|U_{|\set{S}|-1})\big\}, \IEEEeqnarraynumspace
\end{IEEEeqnarray}
where the maximization is over all random tuples $U_1, \ldots, U_{|\set{S}|-1},X, Y_{j_1}, \ldots, Y_{j_{|\set{S}|}}$ that satisfy \eqref{eq:markov_channel} and  \eqref{eq:markovS}. 

Notice that $\C_{\{k\}}$ is simply the point-to-point capacity to receiver~$k$ and we will abbreviate it as $\C_k$.

\section{Coding Schemes and Lower Bounds on the (Global) Capacity-Memory Tradeoff}\label{sec:lower}

\subsection{{The} Local Caching Gain} 

The simplest way to use  receiver cache memories is to store  the same information at each and every receiver. This allows the receivers to retrieve this information locally, without transmission    over the BC. Further global caching gains are not possible under this caching strategy. 


Applying the described caching strategy to only a part of the cache memory that is of size $\Delta \geq 0$, while allowing a smarter use of the remaining memory,  leads to the following proposition, see also \cite[Proposition~1]{WiggerTimoShamai16}. 
\begin{proposition}[Local caching gain]\label{prop:local}
	For all $\Delta>0$ and $\M_1, \ldots, \M_K \geq 0$:
	\begin{equation}
	\C(\M_1+ \Delta, \ldots, \M_K+\Delta)  \geq \C(\M_1,\ldots, \M_K)+ \frac{\Delta}{D}.
	\end{equation} 
	As a consequence, 
	for all $\Delta_{\textnormal{total}}>0$ and $\M\geq 0$:
	\begin{equation}
	\C^\star (\M +\Delta_{\textnormal{total}})  \geq \C^\star(\M)+ \frac{\Delta_{\textnormal{total}}}{K\cdot D}.
	\end{equation} 
\end{proposition}

We will see that in some regimes this lower bound is tight.

\subsection{Superposition Piggyback-Coding}\label{sec:gen_piggyback}
We generalize the piggyback coding for erasure BCs in \cite{saeeditimowigger-IT, saeediwiggertimo-turbo} to general degraded BCs by introducing superposition coding. The idea is to piggyback {information of} multiple stronger receivers on {that of a} single weak receiver. This scheme is {efficient} when a receiver is strictly weaker than the others. Specifically, we assume 
\begin{equation}\label{eq:weaker}
I(U_1^\star; Y_1) < I(U_1^\star; Y_{k}), \quad k\in\{2,\ldots, K\},
\end{equation}
where $(U_1^\star, \ldots, U_{K-1}^\star,X^\star)$ is a random $K$-tuple that achieves the symmetric-capacity $\C_\K$, i.e., it is a solution to the optimization problem in \eqref{eq:common_msg}.


\textit{Preliminaries:} 
 Let $\epsilon>0$ be arbitrary small, and define the rates
\begin{subequations}\label{eq:RAB}
	\begin{IEEEeqnarray}{rCl}
		R^{(\A)} &: = & \C_\K- \epsilon, \label{eq:RA}\\ 
		R^{(\B)}& :=&  \frac{1}{K-1}\big(I(U_1^\star;Y_2)- I(U_1^\star;Y_1)\big).  \label{eq:RB} \IEEEeqnarraynumspace
	\end{IEEEeqnarray}
\end{subequations}
The RHS of \eqref{eq:RB} is positive  by \eqref{eq:weaker}.

Split each message $W_d$, $d\in\{1,\ldots, D\}$, into two parts:
\begin{equation*}
W_d =\big(W_d^{(\A)}, W_{d}^{(\B)}\big),
\end{equation*}
where $W_d^{(\A)}$ and $W_d^{(\B)}$ are of rates $R^{(\A)}$ and $R^{(\B)}$, and thus the total message rate is 
\begin{IEEEeqnarray}{rCl}
	R&=& R^{(\A)}+ R^{(\B) }.
\end{IEEEeqnarray}

Define 
\begin{equation}\label{eq:M1single}
\M_1^{\textsf{single}}:=D \cdot R^{(\B)} =  \frac{D}{K-1}\big(I(U_1^\star;Y_2)- I(U_1^\star;Y_1)\big),
\end{equation}
and allocate the cache size 
\begin{subequations}\label{eq:caches_superpos}
	\begin{equation}\label{eq:M11}
	\M_1= \M_1^{\textsf{single}}
	\end{equation}
	to receiver $1$ and zero cache size to the other receivers
	\begin{equation}
	\M_2= \ldots = \M_K=0.
	\end{equation}
\end{subequations}


\textit{Placement Phase:} 
Store  $W_1^{(\B)},\ldots, W_D^{(\B)}$ in the cache memory of receiver~$1$. This is possible by \eqref{eq:M11}.

\textit{Delivery Phase:} For the transmission in the delivery phase,
 construct a $K$-level superposition code   $\set{C}$ with a cloud center of rate~$R^{(\A)}+(K-1)R^{(\B)}$ and  satellites of rates~$R^{(\A)}$ in Levels $2, \ldots, K$. 
For the code construction, use a
probability distribution
\begin{equation*}
P_{U_1^\star}\cdot P_{U_2^\star|U_1^\star}  \ldots P_{U_{K-1}^\star|U_{K-2}^\star} \cdot P_{X^\star|U_{K-1}^\star}
\end{equation*}
that achieves $\C_\K$.

It will be convenient to {arrange the}  codewords in the cloud center in an array with $\lfloor 2^{nR^{(\A)}} \rfloor$ columns and $(\lfloor 2^{nR^{(\B)}}\rfloor)^{K-1}$ rows. The columns are  used to encode message $W_{d_1}^{(\A)}$ and the rows to encode the message tuple
 \begin{figure}[t!]
	\begin{center}
		\includegraphics[width=0.9\textwidth]{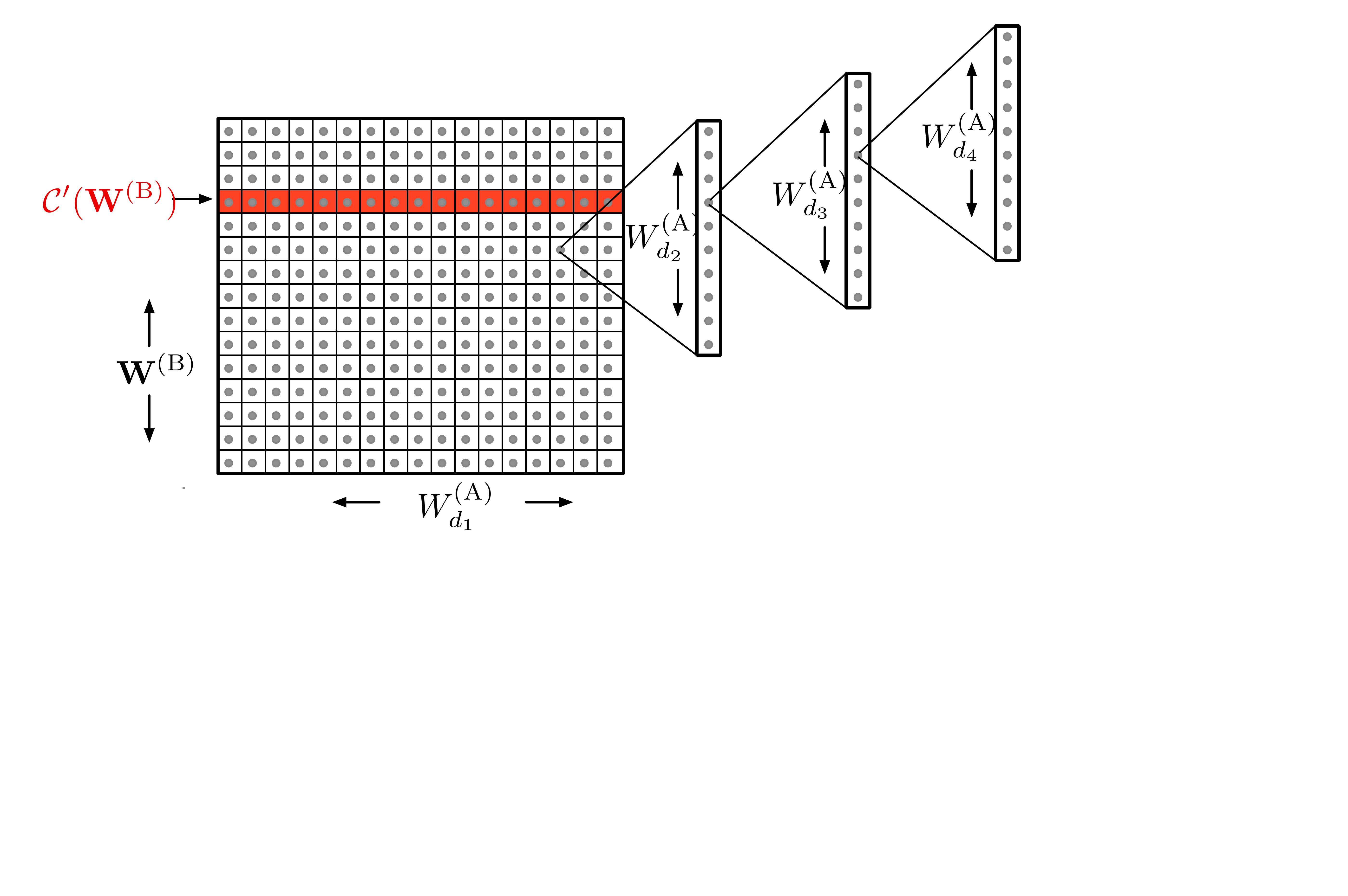}
		\vspace{-3.6cm}
		\caption{Codebook $\set{C}$ for superposition piggyback coding for $K=4$. Each dot represents a codeword.} 
		\label{fig:piggyback}
	\end{center}
\end{figure}
\begin{equation}\label{eq:msg_tuple}
\mathbf{W}^{(\B)}:= \big( W_{d_2}^{(\B)}, \ldots, W_{d_{K-1}}^{(\B)}, W_{d_K}^{(\B)}\big).
\end{equation}
The $k$-th  level satellite is   used to encode message $W_{d_k}^{(\A)}$, for $k\in\{2,\ldots, K\}$.  See Figure~\ref{fig:piggyback} for an illustration of the code construction.

Let $u_1^n(w_{1,\textnormal{column}}, w_{1, \textnormal{row}})$ denote the cloud-center codeword of $\set{C}$ in column $w_{1,\textnormal{column}}$ and row $w_{1, \textnormal{row}}$. Similarly, let $x^n( w_{1,\textnormal{column}}, w_{1, \textnormal{row}}; w_2; w_3; \ldots; w_K)$ denote the Level-$K$ satellite codeword of $\set{C}$ that corresponds to the cloud center codeword $u_!^n(w_{1,\textnormal{column}}, w_{1, \textnormal{row}})$  and to the $w_2$-th, $w_3$-th ,$w_4$-th, etc. satellite codewords in Levels $2,3, 4,\ldots $.

The transmitter  chooses and sends the codeword
\[
x^{ n}\Big( {W_{d_1}^{(\A)},\mathbf{W}^{(\B)} }; \ W_{d_2}^{(\A)};\  W_{d_3}^{(\A)}; \ \ldots \ ; W_{d_K}^{(\A)}\Big)
\]
over the channel.

\textit{Decoding:} 
Receiver~$k\in\{2, \ldots, K\}$,  decodes all messages in Levels $1,\ldots, k$. Recall that its desired message parts $W_{d_k}^{(\A)}$ and $W_{d_k}^{(\B)}$ are encoded in levels~$k$ and $1$ (i.e., the cloud center), respectively. 

Receiver~$1$ only has to decode $W_{d_1}^{(\A)}$, because it can retrieve $W_{d_1}^{(\B)}$ directly from its cache memory. To decode  $W_{d_1}^{(\A)}$ it performs the following steps: 
\begin{enumerate}
	\item It  retrieves the message-tuple   $\mathbf{W}^{(\B)}$
	 from its cache memory.  
	\item It forms the subcodebook $\set{C}'(\mathbf{W}^{(\B)})\subseteq\set{C}$ that contains all level-$1$ codewords that are ``compatible" with the retrieved tuple $\mathbf{W}^{(\B)}$: 
	\begin{equation}
	\set{C}'\big(\mathbf{W}^{(\B)}\big): = \big\{u_1^{n}\big(w, \mathbf{W}^{(\B)}\big) \big\}_{w=1}^{\big\lfloor2^{nR^{(\A)}}\big\rfloor}.
	\end{equation} 
	Figure~\ref{fig:piggyback} illustrates such a subcodebook  in red.
	\item It decodes its desired message $W_{d_1}^{(\A)}$ using an optimal decoding rule for  subcodebook $\set{C}'(\mathbf{W}^{(\B)})$.
\end{enumerate}


\textit{Error Analysis:} 
Each receiver~$k\in\{2,\ldots, K\}$  reliably decodes  messages $(W_{d_1}^{(\A)}, W_{d_2}^{(\B)}, \ldots, W_{d_K}^{(\B)})$ and  $W_{d_2}^{(\A)}, \ldots, W_{d_k}^{(\A)}$ if the following inequalities hold:
\begin{subequations}\label{eq:constraints}
	\begin{align}
	kR^{(\A)}+(K-1)\cdot R^{(\B)}& <  I(U_k^\star;Y_k),\\
	(k-\ell) \cdot R^{(\A)} &< I(U_k^\star;Y_k|U_\ell^\star),\quad \ell\in\{1,\ldots, k-1\}.
	\end{align}
\end{subequations}
One can verify that for  degraded BCs the choice of $R^{(\A)}$ and $R^{(\B)}$ in \eqref{eq:RAB} satisfies  the constraints in \eqref{eq:constraints}.

Finally, receiver~1 can decode with arbitrarily small probability of error because  subcodebook $\set{C}'(\mathbf{W}^{(\B)})$ contains $\lfloor2^{nR^{(\A)}}\rfloor$  codewords  that are generated i.i.d. according to $P_{U_1^\star}$ and because
\[R^{(\A)} < I(U_1^\star;Y_1).\]

Letting $\epsilon \to 0$, we obtain the following result.
\begin{theorem} \label{thm:weak}
	Under cache assignment \eqref{eq:caches_superpos}, we have
	\begin{equation}\label{eq:lower_Ga}
	\C(\M_1,\ldots, \M_K)\geq \C_\K + \frac{\M_1}{D}. 
	\end{equation}
\end{theorem} 

\begin{remark}
	Since receivers can always choose to ignore their cache memories, and because the superposition piggyback coding scheme can be  time- and memory-shared with a no-caching scheme, Theorem~\ref{thm:weak} remains valid for  all
	\begin{IEEEeqnarray}{rCl}
		0 \leq 	\M_1 \leq \M_1^{\mathsf{single}},\\
		\M_2, \ldots, \M_K \geq 0.
	\end{IEEEeqnarray} 
	
\end{remark}

We will see in Corollary~\ref{cor:cache1_exact} ahead, that  \eqref{eq:lower_Ga} holds with equality for all $0\leq \M_1\leq \M_1^{\mathsf{single}}$ provided that  $\M_2=\ldots= \M_K=0$. 

The RHS of \eqref{eq:lower_Ga} coincides  with the capacity-memory tradeoff of a scenario where  each and every receiver has access to receiver~1's cache memory. Superposition piggyback coding  can thus be viewed as a coding  technique that virtually provides all stronger receivers access to the weakest receiver's cache memory.  This is achieved  by transmitting the extra-message tuple $\mathbf{W}^{(\B)}$ in the cloud center and by adapting the decoding at receiver~1  in a way that this additional communication does not influence its decoding performance. 

\subsection{Generalized Coded-Caching}\label{sec:multipg}
We generalize the coded-caching scheme of \cite{maddahali_niesen_2014-1} to noisy BCs with unequal channel conditions and  to receivers with unequal cache sizes. 

We first explain the scheme for a simple special case.

\subsubsection{Special Case $K=2$ and $t=1$}
Fix an input distribution $P_X$ and a small $\epsilon>0$, and define the rates 
\begin{IEEEeqnarray}{rCl}
	R^{(\A)}&=&I(X;Y_1)-\epsilon\\
	R^{(\B)}&=&I(X;Y_2)-\epsilon.
\end{IEEEeqnarray}
Notice that by the degradedness of the BC: 
\begin{equation}
R^{(\B)} \geq R^{(\A)}.
\end{equation}

Fix a blocklength $n$ and generate a random codebook 
\begin{equation}
\set{C}:= \big\{ x^n(j)\big\}_{j=1}^{\lfloor 2^{nR^{(\B)}} \rfloor }
\end{equation} by choosing all entries i.i.d. according to $P_X$. The  codebook $\set{C}$ is revealed to all terminals of the network. 

Allocate cache memories
\begin{subequations}\label{eq:caches}
	\begin{IEEEeqnarray}{rCl}
		\M_1 &= & D\cdot R^{(\B)}= D \cdot (I(X;Y_2)- \epsilon),\\
		\M_2 & = & D \cdot R^{(\A)}= D \cdot (I(X;Y_1)- \epsilon),
	\end{IEEEeqnarray}
\end{subequations}
to receivers~1 and 2, respectively. 

Split  each message $W_{d}$, for $d\in\{1,\ldots, D\}$, into two parts: 
\[W_d =\big(W_d^{(\A)}, W_{d}^{(\B)}\big),
\] 
which are of rates $R^{(\A)}$ and $R^{(\B)}$, respectively.

In  the caching phase, the transmitter stores 
messages 
\[
W_{1}^{(\B)}, \ldots, W_{D}^{(\B)}
\] in  receiver~1's cache memory and messages 
\[W_{1}^{(\A)}, \ldots, W_{D}^{(\A)}\] in  receiver~2's cache memory. This is possible given the  cache assignment in \eqref{eq:caches}.

In the delivery phase the  transmitter uses codebook $\set{C}$ to send   the XOR message\footnote{Recall that in Section \ref{sec:notation} we defined the XOR operation $\bar{\oplus}$ over the binary representations  of the two messages of same length.}
\begin{equation}
\label{eq:XOR_msg}
W_{d_1}^{(\A)} \bar{\oplus} W_{d_2}^{(\B)} 
\end{equation}
to both receivers using the codeword 
\[x^n\Big(W_{d_1}^{(\A)} \bar{\oplus} W_{d_2}^{(\B)} \Big).
\]

Receiver~2 decodes the XOR-message, and XORs the decoded message with $W_{d_1}^{(\A)}$, which  it has stored in its cache memory. It then combines this guess of $W_{d_2}^{(\B)}$ with the message $W_{d_2}^{(\A)}$ from its cache memory.

Receiver~1 
performs joint cache-channel decoding where it can exploit that it has more cache memory than receiver~2. Specifically, it retrieves $W_{d_2}^{(\B)}$ from its cache memory, and extracts a subcodebook~$\set{C}'(W_{d_2}^{(\B)})\subseteq \set{C}$ containing all  codewords that are compatible with $W_{d_2}^{(\B)}$: 
\begin{equation}
\set{C}' \big(W_{d_2}^{(\B)}\big): = \Big\{ w \; \bar{\oplus} \; W_{d_2}^{(\B)} \Big\}_{w=1}^{\big\lfloor 2^{nR^{(\A)}}\big\rfloor}
\end{equation}
Note that  subcodebook $\set{C}'(W_{d_2}^{(\B)})$ is of rate $R^{(\A)}$ which is smaller than the rate $R^{(\B)}$ of the original codebook $\set{C}$.

Receiver~1 then decodes the XOR message in \eqref{eq:XOR_msg} using an optimal decoding rule for this subcodebook~$\set{C}'(W_{d_2}^{(\B)})$, and it XORs the decoded message with  $W_{d_2}^{(\B)}$, which it has stored in its cache memory. It then combines the resulting guess of $W_{d_1}^{(\A)}$ with the message  $W_{d_1}^{(\B)}$ from its cache memory. 

Since both receivers correctly guess their desired messages $W_{d_1}$ and  $W_{d_2}$ whenever they successfully decode the XOR-message in \eqref{eq:XOR_msg}, and since  the rate $R^{(\B)}$ of the original codebook~$\set{C}$ satisfies
\begin{equation}
R^{(\B)} < I(X;Y_2),
\end{equation}and the rate of $R^{(\A)}$ of  the subcodebook  $\set{C}'(W_{d_2}^{(\B)})$ satisfies 
\begin{equation}
R^{(\A)} < I(X;Y_1), 
\end{equation}
the probability of decoding error at both receivers tends to 0 as the blocklength $n$ tends to infinity. 

Letting $\epsilon \to 0$, we conclude that for $K=2$ the rate-memory triple 
\begin{IEEEeqnarray}{rCl}
	R & = & I(X;Y_1) +I(X;Y_2),\nonumber\\
	M_1& = & I(X;Y_2), \nonumber \\
	M_2& = & I(X;Y_1),\nonumber
\end{IEEEeqnarray}
is achievable.

Notice that the weaker receiver~1 is assigned a larger cache memory than the stronger receiver~2: 
\begin{equation}
\M_1\geq \M_2.
\end{equation}
The described scheme can also be applied with a uniform cache assignment $\M_1=\M_2 = D\cdot R^{(\A)}$,  however at the cost of  a decreased achievable rate $R=2 \cdot I(X;Y_1)$. In fact, assigning a larger cache memory $\M_1$ to receiver~1 allows to transmit more information to receiver~2 during the communication to receiver~1. 

\subsubsection{General Scheme}
We will need the following definitions.
Let for each $t\in\K$  
\begin{subequations}\label{eq:Gell}
	\begin{equation}
	\set{G}_1^{( t)}, \ldots, \set{G}_{ { K \choose  t}}^{( t)}
	\end{equation}
	denote all unordered size-$t$ subsets of~$\K$. Define their complements as: 
	\begin{equation}
	\set{G}_\ell^{( t),c}:= \K \backslash \set{G}_\ell^{( t)}, \qquad \ell\in\Big\{1, \ldots, {K \choose t}\Big\}.
	\end{equation}
\end{subequations}
Pick a small number $\epsilon>0$ and an input distribution  $P_X$. Pick further a parameter $t\in\{1,\ldots, K-1\}$, and assign the following cache size  to receiver~$k\in \K$: 
\begin{IEEEeqnarray}{rCl}\label{eq:achM}
	\M_k^{(t)}&:= &{D} \cdot  \frac{\sum_{\big\{ \ell \colon \;  k \in  \set{G}_\ell^{({t})}\big\}} \prod_{k'\in\set{G}_{\ell}^{(t),\c}} I(X;Y_{k'})}{\sum_{j=1}^{{K \choose t+1}} \prod_{k'\in{\set{G}}_{j}^{(t+1),\c}} I(X;Y_{k'})}  \nonumber \\
	& & - D {K-1 \choose t-1} \epsilon.
	\label{eq:Mach}
\end{IEEEeqnarray}
Notice that 
\begin{IEEEeqnarray*}{rCl}
	\M_{1}^{(t)} \leq \M_{2}^{(t)} \leq \cdots \leq \M_{K}^{(t)},  &\qquad& t\in\{1,\ldots, K-1\}, 
	\end{IEEEeqnarray*}
so a larger cache memory is assigned the weaker a receiver is.
%

Split each message  $W_d$ into  ${K \choose t}$ independent submessages: 
\[
W_d =\left\{ W_{d, \set{G}_\ell^{(t)}}\colon \quad\ell=1,\ldots, {K \choose t}\right\},
\]
where each submessage $W_{d, \set{G}_\ell^{(t)}}$ is  of rate
\begin{equation}\label{eq:Rell}
R_{\set{G}_\ell^{(t)}} := \frac{\prod_{k\in\set{G}_{\ell}^{(t),\c}} I(X;Y_k)}{\sum_{j=1}^{K\choose t+1} \prod_{k\in \set{G}_j^{(t+1),\c} }I(X;Y_k)} - \epsilon.
\end{equation}
The total message rate is thus
\begin{equation}\label{eq:point3}
R:=\sum_{\ell=1}^{K \choose t}  R_{\set{G}_\ell^{(t)}} =\frac{\sum_{\ell=1}^{{K \choose t}}  \prod_{k\in{\set{G}}_{\ell}^{({t}),\c}} I(X;Y_k)}{\sum_{j=1}^{K\choose t+1} \prod_{k\in \set{G}_j^{(t+1),\c} }I(X;Y_k)} - {K \choose t} \epsilon.
\end{equation}

Notice that when  $t=K-1$ the denominator of \eqref{eq:achM}, \eqref{eq:Rell}, and \eqref{eq:point3} all equal $1$.

\textit{Placement Phase:}
For each $d\in\{1,\ldots, D\}$, store the tuple
\begin{equation}\label{eq:cached}
\left\{  W_{d, \set{G}_{\ell}^{({t})}}  \colon  \; k \in  \set{G}_\ell^{({t})} \right\}.
\end{equation}
in the cache memory of receiver~$k\in\K$. This is possible by  \eqref{eq:Rell} and the cache assignment in \eqref{eq:Mach}.


\textit{Delivery Phase:}
Transmission in the delivery phase takes place in ${K \choose t+1}$ subphases. 

A given subphase~$j\in\big\{1,\ldots, {K \choose t+1}\big\}$ is of length
\begin{equation}\label{eq:nj}
n_j := \left \lfloor n \cdot \frac{\prod_{k\in{\set{G}}_j^{(t+1),\c}} I(X;Y_k)}{\sum_{j'=1}^{K\choose t+1} \prod_{k\in \set{G}_{j'}^{(t+1),\c} }I(X;Y_k)}\right \rfloor,
\end{equation}
and is used to transmit messages 
\begin{equation}\label{eq:msgs_periodj}
\Big\{W_{d_k, \set{G}_{j}^{(t+1)} \backslash \{k\}}\Big\}_{k \in \set{G}_j^{(t+1)}}
\end{equation}
to the intended receivers in $\set{G}_j^{(t+1)}$. For this purpose, the transmitter creates the 
 XOR message %
	\begin{equation}\label{eq:XORSj}
	{W}_{\textnormal{XOR},\set{G}_j^{(t+1)}} = \overline{\bigoplus}_{k \in \set{G}_j^{(t+1)}} {W}_{d_{k},\set{G}_{j+1}^{(t)}\backslash\{k\}},
	\end{equation}
	which is of rate
\begin{IEEEeqnarray}{rCl}\label{eq:Rj}
\lefteqn{R_{\textnormal{XOR},\set{G}_j^{(t+1)}} 
	:=  \max_{\set{G}_{\ell}^{(t)} \subseteq \set{G}_j^{(t+1)}} R_{\set{G}_\ell^{(t)}}}  \nonumber \\
	& = & \left(\max_{k'\in\set{G}_j^{(t+1)}}I(X;Y_{k'})\right) \cdot \frac{\prod_{k\in{\set{G}}_{j}^{(t+1),\c} } I(X;Y_k)}{\sum_{j=1}^{K\choose t+1} \prod_{k\in \set{G}_j^{(t+1),\c} }I(X;Y_k)} -\epsilon,\nonumber \hspace{3mm}\\
\end{IEEEeqnarray}
and generates a codebook 
\begin{equation}
\set{C}_j = \bigg\{  x_j^{n_j}( w)  \colon \; w=1, \ldots, \Big\lfloor2^{n R_{\textnormal{XOR},\set{G}_j^{(t+1)}}} \Big \rfloor \bigg\},
\end{equation}
by drawing all entries i.i.d. according to  $P_X$. 

The transmitter then sends the codeword 
\begin{equation}
 x_j^{n_j}\left({W}_{\textnormal{XOR},\set{G}_j^{(t+1)}} \right)  
\end{equation}
over the channel.

We now describe the decoding. Each receiver~$k\in\K$ can retrieve messages 
\begin{IEEEeqnarray}{rCl}\label{eq:cachedk}
	\Big\{  W_{d_k, \set{G}_{\ell}^{({t})}}  \colon \;   k \in  \set{G}_\ell^{({t})}\Big\}
\end{IEEEeqnarray}
directly from its cache, see \eqref{eq:cached}, and thus only needs to decode messages
\begin{IEEEeqnarray}{rCl}\label{eq:decodedk}
	\Big\{  W_{d_k, \set{G}_{\ell}^{({t})}} \colon \; k \notin  \set{G}_\ell^{({t})}\Big\}.
\end{IEEEeqnarray}
For each $j\in\{1,\ldots, {K\choose t+1}\}$ and  $k\in\set{G}_j^{(t+1)}$, receiver~$k$ decodes message $W_{d_k, \set{G}_j^{(t+1)} \backslash \{k\}}$ from its subphase-$j$ outputs
\begin{IEEEeqnarray*}{rCl}
	Y_{k,j}^{n_j}&:=&\big(Y_{k,\sum_{j'=1}^{j-1} n_{j'}+1}, \ldots, Y_{k,\sum_{j'=1}^{j} n_{j'}}\big).
\end{IEEEeqnarray*} 
Specifically, 	with the messages stored in its cache memory,  it forms the XOR message 
	\begin{equation}\label{eq:retrieve}
W_{\textnormal{XOR}, j, k}:=	\displaystyle  \overline{ \bigoplus}_{{k' \in \set{G}_j^{(t+1)} \backslash \{k\}} } \;
	 W_{d_{k'}, \set{G}_{j}^{(t+1)} \backslash \{k'\}},
	\end{equation}
 and it extracts a subcodebook $\set{C}_{j,k}'( W_{\textnormal{XOR}, j, k})$ from $\set{C}_j$ that contains  all codewords that are compatible with  $W_{\textnormal{XOR}, j, k}$: 	
	\begin{IEEEeqnarray*}{rCl}
\set{C}_{j,k}'( W_{\textnormal{XOR}, j, k}) &:=& \Big\{ x_j^{n_j}\big( w \;  \bar{\oplus} \; W_{\textnormal{XOR}, j, k}  \big) \colon  
\quad w=1,\ldots, \Big\lfloor 2^{n R_{\set{G}_{j}^{(t+1)}\backslash\{k\}}}\Big\rfloor \Big\}.
	\end{IEEEeqnarray*}
	It then decodes the XOR message  ${W}_{\textnormal{XOR},\set{G}_j^{(t+1)}}$ by applying an optimal decoding rule for subcodebook $\set{C}_{j,k}'( W_{\textnormal{XOR}, j, k}) $ to the subphase-$j$ outputs $Y_{k,j}^{n_j}$, and XORs the resulting guess $\hat{W}_{\textnormal{XOR},\set{G}_j^{(t+1)}}$ with   $W_{\textnormal{XOR}, j, k}$ to obtain 
	\begin{equation}
	\hat{W}_{d_k, \set{G}_j^{(t+1)} \backslash \{k\}} = \hat{{W}}_{\textnormal{XOR},\set{G}_j^{(t+1)}}\;  \bar{\oplus} \; W_{\textnormal{XOR}, j, k}.
	\end{equation}
	
	After the last sub-phase ${ K \choose t+1}$, each receiver~$k\in\K$ has decoded all its missing messages in \eqref{eq:decodedk}, and can thus produce a final guess of message $W_{d_k}$.

\textit{Error Analysis:}
If each  XOR-message ${W}_{\textnormal{XOR},\set{G}_j^{(t+1)}}$ is decoded correctly by all its intended receivers in $\set{G}_j^{(t+1)}$, $j=1,\ldots,{K \choose t+1}$, then all receivers~$1, \ldots, K$ produce the correct estimate of their desired messages $W_{d_1}, \ldots, W_{d_K}$.

The probability that   receiver~$k\in\set{G}_j^{(t+1)}$ wrongly decodes the XOR message ${W}_{\textnormal{XOR},\set{G}_j^{(t+1)}}$  tends to 0 as $n$ (and thus $n_j$) $\to\infty$ because 
the  rate of the subcodebook  $\set{C}_{j,k}'$ satisfies
\[
\varlimsup_{n\to \infty} \frac{n}{n_j}\cdot R_{  \set{G}_{j}^{(t+1)} \backslash \{k\}}< I(X;Y_k),
\]
see \eqref{eq:Rell} and \eqref{eq:nj}.

By letting $\epsilon \to 0$, we  conclude the following result. 
\begin{theorem}\label{thm:multipg} 
	Fix a $t\in\{1,\ldots, K-1\}$ and an input distribution $P_X$, and consider the corresponding cache assignment in \eqref{eq:achM}. 	
	Then, 
	\begin{IEEEeqnarray}{rCl}\label{eq:ineq_generalized}
		\C\Big(\M_1^{(t)}, \ldots, \M_K^{(t)}\Big) &\geq & R^{(t)},
		\IEEEeqnarraynumspace
	\end{IEEEeqnarray}
	where $R^{(t)}$ is calculated from $P_X$ as described in \eqref{eq:point3}.	
\end{theorem}

As we will see in Corollary~\ref{cor:generalizedtight}, the Inequality in \eqref{eq:ineq_generalized} holds with equality for $t=K-1$.

%
%
\subsection{Lower Bound on $\C^\star(\M)$}

Proposition~\ref{prop:local} and Theorems~\ref{thm:weak} and \ref{thm:multipg} readily yield a lower bound on $\C^\star(\M)$. As we will see in Corollary~\ref{cor:exact} ahead, this lower bound is exact in the regimes of small and large total cache size $\M$. 

 Let 
\begin{subequations}\label{eq:RMtotal}
	\begin{equation}
	R^{(0)}:= \C_\K, \qquad  \quad \M^{(0)}:=0,
	\end{equation}
	and 
	\begin{equation}\label{eq:Msingle}
	R^{{\mathsf{single}}}:= \C_\K+\frac{\M^{{\mathsf{single}}}}{D}, \qquad  \quad \M^{\mathsf{single}}:=\M_1^{\mathsf{single}},
	\end{equation}
	where $ \C_\K$ is defined in \eqref{eq:C0op} and $\M_1^{\mathsf{single}}$ is defined in \eqref{eq:M1single}. Also, for given $P_X$, recall $\M^{(t)}$ and $R^{(t)}$ from \eqref{eq:achM} and \eqref{eq:point3}, and define for $t\in\{1,\ldots, K-1\}$:
	\begin{equation}
	\M^{(t)}:= \sum_{k=1}^K \M_k^{(t)}.
	\end{equation}
\end{subequations}

\begin{proposition}\label{thm:lower_global}
	For any $P_X$, all rate-memory pairs in \eqref{eq:RMtotal} are achievable.  By time- and memory-sharing arguments,  the upper-convex envelope of all these rate-memory pairs lower bounds $\C^\star(\M)$:
	\begin{IEEEeqnarray}{rCl}
		\lefteqn{\C^\star(\M) \geq \textnormal{upp hull}\Big\{\big(R^{(0)}, \M^{(0)}\big), \ \big(R^{\mathsf{single}}, \M^{\mathsf{single}}
		\big) ,
			 }\nonumber \\
		&&  \hspace{1.3cm} \  \bigcup_{P_X} \Big\{ \big(R^{(1)}, \M^{(1)}\big), \ldots,  \ \big(R^{(K-1)}, \M^{(K-1)}\big) \Big\}
		\Big\}.\IEEEeqnarraynumspace
	\end{IEEEeqnarray} 
\end{proposition}

Notice that for any $P_X$:
\begin{equation}
\M^{(0)} \leq \M^{\mathsf{single}}\leq \M^{(1)} \leq \cdots \leq  \M^{(K-1)}
\end{equation}
and
\begin{equation}
R^{(0)} \leq R^{\mathsf{single}}\leq R^{(1)} \leq \cdots \leq  R^{(K-1)}.
\end{equation}

\section{Upper Bounds and Exact Results on Global Capacity-Memory Tradeoff} \label{sec:upper}

\subsection{Results on $\C(\M_1, \ldots, \M_K)$}
The  upper bound  is formulated in terms of the following parameters. For each receiver set $\set{S}$ as in \eqref{Aylin}, define 
 \begin{subequations}\label{eq:values_of_alpha}
		\begin{IEEEeqnarray}{rCl}
			\alpha_{\set{S},1}^\star &:=& \frac{\M_{j_1}}{D}
		\end{IEEEeqnarray}
		and for ${k}\in\{2,\ldots,|\set{S}|\}$:
		\begin{IEEEeqnarray}{rCl}
			\alpha_{\set{S},{k}}^\star&:=& \min\Bigg\{\frac{{\sum_{i =1}^{k} \M_{{j_i}}}}{D-{k}+1}, \; {\frac{1}{|\set{S}|-k+1}{\Bigg( \frac{|\set{S}|}{D} \sum_{i=1}^{|\set{S}|} \M_{j_i} - \sum_{i=1}^{{k}-1} \alpha_{\set{S},{i}} \Bigg)}} \Bigg\}.\IEEEeqnarraynumspace
		\end{IEEEeqnarray}
	\end{subequations}

\begin{theorem}\label{thm:upper_new}There exist random variables $X, Y_1, \ldots, Y_K$ and for \emph{every} receiver set $\set{S}$ as in \eqref{Aylin} random variables $\{U_{\set{S},1}, \ldots, U_{\set{S},{|\set{S}|-1}} \}$  so that  the channel law \eqref{eq:markov_channel} and  the Markov chain 
	\begin{equation}\label{eq:markovsetS}
U_{\set{S},1} - U_{\set{S},2} -U_{\set{S},|\set{S}|}- \cdots -  U_{\set{S},|\set{S}|-1} -X - \big(Y_1, \ldots,Y_K \big)
\end{equation}  
	hold   
	and so that  for each $\set{S}$:
	\begin{subequations}\label{eq:d_inequa10}
		\begin{IEEEeqnarray}{rCl}
			\C(\M_1,\ldots, \M_K)&\leq&I\big(U_{\set{S},1};Y_{{j_1}} \big)+ \alpha_{\set{S},1}^\star ,\\
			\C(\M_1,\ldots, \M_K)&\leq&I\big(U_{\set{S},k};Y_{{j_k}} | U_{\set{S},k-1})+ \alpha_{\set{S},k}^\star,  \quad \forall k\in\{2,\ldots, {|\set{S}|-1}\},\IEEEeqnarraynumspace \\
			\C(\M_1,\ldots, \M_K)&\leq&I\big(X;Y_{{j_{|\set{S}|}}} | U_{\set{S},|\set{S}|-1})+ \alpha_{\set{S},|\set{S}|}^\star.
		\end{IEEEeqnarray}
	\end{subequations}
	\end{theorem}
\begin{IEEEproof}
	See Appendix~\ref{sec:upperbound}.
\end{IEEEproof}


	Without cache memories, $\M_1=\ldots =\M_K=0$, the parameters $\alpha_{\set{S},1}^\star, \ldots, \alpha_{\set{S},|\set{S}|}^\star$ equal 0  for all $\set{S}\subseteq \{1,\ldots,K\}$,  and the upper bound in Theorem \ref{thm:upper_new} recovers the  exact capacity-memory tradeoff  $\C_\K$ in \eqref{eq:common_msg}. 

\medskip
 
	The upper bound in Theorem~\ref{thm:upper_new} is asymmetric in  the different cache sizes $\M_1,\M_2,\ldots,\M_K$, because the parameters $\alpha_{\set{S}, j_i}^\star$ are not symmetric. In fact, increasing the cache memories at weaker receivers generally increases the upper bound more than increasing the cache memories at stronger receivers.


\medskip

The converse in Theorem \ref{thm:upper_new} is weakened if  constraints~\eqref{eq:d_inequa10} are ignored for certain receiver sets $\set{S}$, or if in these constraints the input/output random variables $X, Y_{j_1}, \ldots, Y_{j_{|\set{S}|}}$  are allowed to depend on the receiver set $\set{S}$. For this latter relaxation, Theorem~\ref{thm:upper_new} results in the following corollary.
 \begin{corollary}\label{cor-equivalent}
Given  cache sizes  $\M_1, \ldots, \M_K\geq 0$,  rate~$R$ is achievable only if for every receiver set $\set{S}\subseteq \K$: \begin{equation}
	\left(R- \alpha_{\set{S},1}^\star, R-\alpha_{\set{S},2}^\star, \ldots, R- \alpha_{\set{S},|\set{S}|}^\star\right)\in\capa_{\set{S}},\label{def:eqthmup}
	\end{equation}
	where $\capa_{\set{S}}$ denotes the capacity region to receivers in $\set{S}$ (ignoring receivers in $\set{K}\backslash \set{S}$) when there are no cache memories.
\end{corollary}

\medskip

\begin{remark}\label{rem:1}
The upper bounds of Theorem~\ref{thm:upper_new} and Corollary~\ref{cor-equivalent} are relaxed when  each $\alpha_{\set{S},k}^\star$ is replaced by $\tilde{\alpha}_{\set{S},k}$, where
\begin{subequations} \label{eq:choice1}
\begin{IEEEeqnarray}{rCl}
\tilde{\alpha}_{\set{S},1}&: =& \frac{\M_{j_1}}{D},\\
\tilde{\alpha}_{\set{S},k} &:=& \frac{|\set{S}|\cdot \sum_{i=1}^{|\set{S}|} \M_{j_i} - \M_{j_1}}{(|\set{S}|-1)D}, \quad k\in\{2\ldots, |\set{S}|\}.\IEEEeqnarraynumspace
\end{IEEEeqnarray}
\end{subequations}
The same holds if each $\alpha_{\set{S},k}^\star$ is replaced by  
\begin{equation}\label{eq:choice2}
{\alpha}_{\set{S},k}': =
\frac{\sum_{i=1}^{|\set{S}|} \M_{j_i}}{D}.
\end{equation}
Replacing in  Corollary~\ref{cor-equivalent} each parameter $\alpha_{\set{S},k}^\star$ by $\alpha_{\set{S},k}' $  recovers the  previous upper bound in \cite[Theorem~9]{saeeditimowigger-IT} and \cite[Theorem 1]{saeediwiggertimo-turbo}. 
\end{remark}
\begin{IEEEproof}
 The proof requires a close inspection of the proof of Theorem~\ref{thm:upper_new} in Appendix~\ref{sec:upperbound}. See  Appendix~\ref{app:proofremark}. 
\end{IEEEproof}

By comparing the new upper bounds with the three achievability results  in the previous Section~\ref{sec:lower}, the exact expression for $\C(\M_1, \ldots, \M_K)$ can be obtained in some special cases. 

The following corollary states that superposition piggyback coding is optimal when only receiver~1 has a cache memory and this cache memory is small. 
\begin{corollary}  \label{cor:cache1_exact}
Under a cache assignment satisfying
	 \begin{equation}\label{eq:single_cach}
	0 \leq  \M_1 \leq \M_1^{\textsf{single}}\quad \textnormal{ and } \quad \M_2=\ldots=\M_K=0,
	 \end{equation} 
	the capacity-memory tradeoff is
	 \[\C(\M_1, 0, \ldots,0)=\C_\K+\frac{\M_1}{D}.
	 \]
	\end{corollary}
	\begin{IEEEproof}
Achievability follows by Theorem \ref{thm:weak}. The converse from Corollary~\ref{cor-equivalent}, where it suffices to consider only the set $\set{S}=\set{K}$. In fact, under \eqref{eq:single_cach}, $\alpha_{\set{K},1}=\ldots= \alpha_{\set{K},K}=\frac{\M_1}{D}.$
	\end{IEEEproof}

The next corollary states that generalized coded caching with parameter $t=K-1$ is optimal under the corresponding cache assignment. Moreover, any extra cache memory that is uniformly distributed over the $K$ receivers only brings local caching gain. 

\begin{proposition}\label{cor:generalizedtight}
For each $k\in\K$, let $\M_k^{\star (K-1)}$ be given  by \eqref{eq:achM} when $P_X$ is chosen as a  maximizer of
\begin{equation}\label{eq:max}
\C_{\textnormal{Avg}}:=\frac{1}{K} \cdot \max_{P_X} \left(\sum_{k=1}^K I(X;Y_k) \right).
\end{equation}
For any $\Delta\geq 0$:
\begin{IEEEeqnarray}{rCl}\label{eq:largeDelta}
\lefteqn{\C\Big(\M_1^{\star (K-1)}+ \Delta, \; \ldots, \; \M_K^{\star (K-1)}+\Delta\Big) } \qquad  \nonumber \\
 &= & \C_{\textnormal{Avg}}+ \frac{\sum_{k=1}^K \M_k^{\star (K-1)}}{K \cdot D} + \frac{\Delta}{D}. \IEEEeqnarraynumspace
\end{IEEEeqnarray}
	\end{proposition}
\begin{IEEEproof}
See Appendix \ref{ap:cor:generalizedtight}.
\end{IEEEproof}

%

\subsection{Results on $\C^\star(\M)$}

Theorem \ref{thm:upper_new} directly yields the following result.
\begin{proposition}\label{thm:newglobal}
	There exist random variables $X, Y_1, \ldots, Y_K$ and for \emph{every} receiver set $\set{S}$ as in \eqref{Aylin}  random variables $\{U_{\set{S},1}, \ldots, U_{\set{S},{|\set{S}|-1}} \}$, such that  \eqref{eq:markov_channel} and  \eqref{eq:markovsetS} hold, and such that \emph{for some} $\M_1,\ldots, \M_K\geq 0$ summing to $\M$ and all $\set{S}$:
	\begin{subequations}\label{eq:cond2}
		\begin{IEEEeqnarray}{rCl}
			\C^\star(\M)&\leq&I\big(U_{\set{S},1};Y_{{j_1}} )+ \alpha_{\set{S},1}^\star, \\
			\C^\star(\M)&\leq&I\big(U_{\set{S},k};Y_{{j_k}} | U_{\set{S},k-1})+ \alpha_{\set{S},k}^\star, \;\; k\in\{2,\ldots, {|\set{S}|-1}\}, \nonumber \\
			\\
			\C^\star(\M)&\leq&I\big(X;Y_{{j_{|\set{S}|}}} | U_{\set{S},|\set{S}|-1})+ \alpha_{\set{S},|\set{S}|}^\star,
		\end{IEEEeqnarray} 
	\end{subequations}
	where $\{\alpha_{\set{S},k}^\star\}$ are defined in~\eqref{eq:values_of_alpha}. 
\end{proposition}
Solving this optimization problem numerically is computationally complex. Simpler, albeit looser, upper bounds can be obtained by either ignoring some of the constraints \eqref{eq:cond2};  by replacing each parameter  $\alpha^\star_{\set{S},k}$ in \eqref{eq:cond2} by $\tilde{\alpha}_{\set{S},k}$ or by $\alpha_{\set{S},k}'$; or by allowing $X,Y_{j_1}, \ldots, Y_{j_\set{S}}$ in \eqref{eq:cond2} to depend on the set $\set{S}$.  

\medskip
The following corollary  presents a simpler bound that is obtained this way.
Recall the definitions in \eqref{eq:Gell}.
\begin{corollary}\label{cor:Global}
	For each $t\in\K$:
	\begin{equation}\label{eq:upg}
	\C^\star(\M) \leq \frac{1}{{K \choose t}}\sum_{\ell=1}^{{K \choose t}}\C_{\set{G}_\ell^{(t)}}+ \frac{t }{K} \cdot\frac{ \M}{ D}.
	\end{equation}
\end{corollary}
\begin{IEEEproof}
	Fix $t\in\K$.
	For each  $\ell=1,\ldots {K \choose t}$, 
	specialize   Corollary~\ref{cor-equivalent} to $\set{S}= \set{G}_{\ell}^{(t)}$ and relax it 
	by replacing each parameter $\alpha_{\set{G}_\ell^{(t)},k}^\star$ by $\alpha_{\set{G}_\ell^{(t)},k}'$.  
	Since $\alpha_{\set{G}_\ell^{(t)},1}'= \ldots= \alpha_{\set{G}_\ell^{(t)},t}'$, we obtain
	\begin{IEEEeqnarray}{rCl}\label{eq:boundG}
		\C^\star(\M) \leq \C_{\set{G}_{\ell}^{(t)}} + \alpha_{\set{G}_\ell^{(t)},1}' = \C_{\set{G}_{\ell}^{(t)}}+ \frac{  \sum_{i\in\set{G}_\ell^{(t)}} \M_{i}}{ D}.
	\end{IEEEeqnarray}
	Now, averaging bound \eqref{eq:boundG} over all indices $\ell=1, \ldots, {K \choose t}$ and upperbounding  the sum $\M_1+ \ldots+ \M_K$ by $\M$ yields the desired result in the corollary.
\end{IEEEproof}


The last result of this section contains  two more simple upper bounds on $\C^\star(\M)$.  For small total cache size $\M$ one of them is achieved by assigning the entire cache memory to the weakest receiver and applying superposition piggyback coding. For large total cache size $\M$  the other is achieved by generalized coded  caching with parameter $t=K-1$, and by first applying the  cache assignment corresponding to this scheme followed by a uniform cache assignment of any remaining cache memory. 
\begin{corollary}\label{cor:exact}
For total cache size $\M\geq 0$:
\begin{equation}\label{eq:exact1}
	\C^\star(\M) \leq \C_\K+ \frac{\M}{  D}
\end{equation}
and 
\begin{equation}\label{eq:ineqularge}
\C^\star(\M) \leq \C_{\textnormal{avg}}+\frac{1}{K} \cdot \frac{\M}{D}.
\end{equation} 
For small cache sizes,  
\begin{equation}
0\leq  \M \leq  \M^{\textsf{ single}},
\end{equation}
\eqref{eq:exact1} holds with equality. 

For large cache sizes, 
\begin{equation}\label{eq:largecache}
\M \geq D \cdot (K-1) \cdot K \cdot \C_{\textnormal{avg}},
\end{equation}
\eqref{eq:ineqularge} holds with equality.
\end{corollary}
\begin{IEEEproof}	Upper bound~\eqref{eq:exact1} follows by specializing Corollary~\ref{cor:Global} to $t=K$. 
	Upper bound~\eqref{eq:ineqularge} is proved as follows. Relax  Theorem~\ref{thm:newglobal} by replacing each parameter $\alpha^\star_{\set{S},k}$  by $\alpha_{\set{S},k}'$ and considering only the  constraints~\eqref{eq:cond2} that correspond  to sets $\set{S}=\{k\}$, for $k\in\K$. Finally, average  the $K$ resulting inequalities and maximize over the input distribution $P_X$.

		The tightness of \eqref{eq:exact1} for $\M\leq \M^{\textsf{single}}$ follows from Theorem~\ref{thm:weak}. 
		The tightness of \eqref{eq:largecache} for 
		$\M \geq D (K-1) K \C_{\textnormal{Avg}}$ 
		 follows from  Proposition~\ref{cor:generalizedtight} because 
		 \[\M_1^{\star(K-1)} + \ldots + \M_K^{\star (K-1)} = D (K-1) K \C_{\textnormal{Avg}}.
		 \]
	\end{IEEEproof}
\medskip 
We remark that for small total cache sizes,  $\C^\star(\M)$ grows as $\frac{\M}{D}$. This corresponds to a \emph{perfect global caching gain}, i.e., the same performance as in a system where each receiver can directly access all cache contents in the network. For large total cache sizes, $\C^\star(\M)$ grows only as $\frac{1}{K} \cdot \frac{\M}{D}$. This corresponds to the local caching gain achieved by Proposition~\ref{prop:local}.

\section{Examples}
\label{sec:examples}
\subsection{Erasure BCs}\label{sec:erasure}

We specialize our results to erasure BCs where at time~$t$ 
receiver~$k$'s output $Y_{k,t}$ equals the channel input $X_t$ with  probability~$1-\delta_k$ and it equals an erasure symbol ``?" with probability~$\delta_k$. The erasure probabilities satisfy: 
\begin{equation}
	1 > \delta_1 \geq \delta_2 \geq \ldots \geq \delta_K \geq 0.
	\end{equation}
	
	For erasure BCs, 
\begin{equation}\label{eq:Csymerasure}
\C_{\set{S}}= \left( \sum_{s\in\set{S}} \frac{1}{1-\delta_s}\right)^{-1}, \qquad \set{S} \subseteq \K.
\end{equation}
Moreover, a  Bernoulli-$1/2$ input distribution $P_X$ maximizes $I(X;Y_k)$ and $I(X;Y_{k}|U)$ simultaneously for all $k\in \K$ and auxiliaries $U$ that form the Markov chain $U - X -Y_k$.  Therefore, Theorem~\ref{thm:upper_new} and Corollary~\ref{cor-equivalent} coincide. Also,  
\begin{IEEEeqnarray}{rCl}
	\C_{\textnormal{avg}} = \frac{1}{K} \sum_{k=1}^K 	\C_{k}= 1- \frac{ \sum_{k=1}^K \delta_k}{K}.
\end{IEEEeqnarray}

\begin{figure}[h!]
	  \input{ErasureBC}
	  \caption{Bounds on $\C^\star(\M)$ for a  4-user Erasure BC with $\delta_1=0.9$, $\delta_2=0.6$, $\delta_3=0.1$, $\delta_4=0.051$.}
	  \label{fig:ErasureBC}
	 \end{figure}
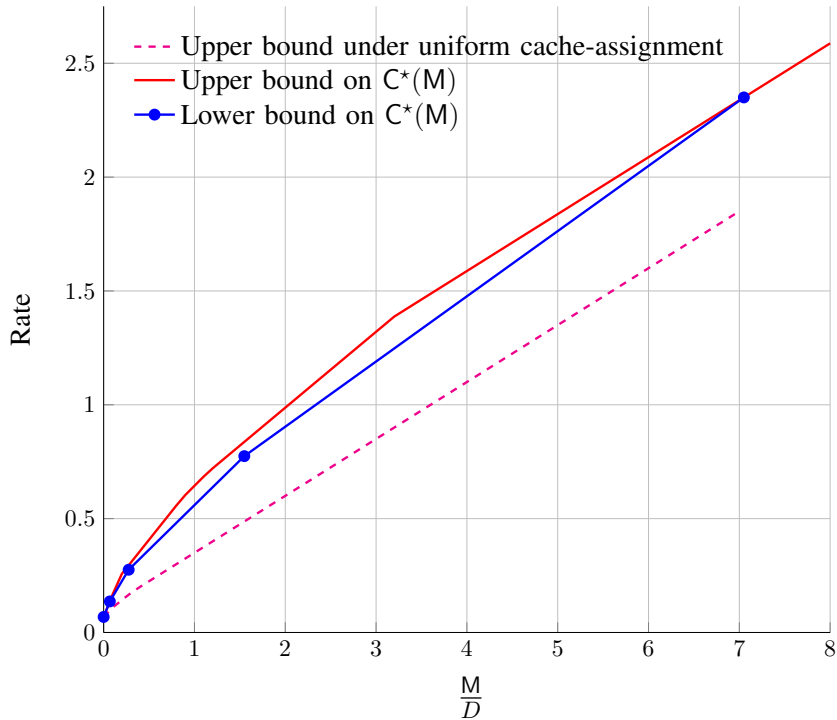

Figure~\ref{fig:ErasureBC}, depicts the  upper and lower bounds on $\C^\star(\M)$ in Propositions~\ref{thm:lower_global} and \ref{thm:newglobal}. 
For comparison, also the upper bound in Theorem~\ref{thm:upper_new} under a uniform cache assignment 
\begin{equation*}
\M_1=\ldots=\M_K=\frac{\M}{K}
\end{equation*} is plotted. This proves numerically  that  a smart allocation of the total cache memory $\M$  significantly increases the global capacity-memory tradeoff of erasure BCs when different receivers have different erasure probabilities. 
	 
Analytically, we can prove that for small total cache size $\M\leq \M^{\textsf{single}}$ any cache assignment that does not allocate all cache memory to the weakest receiver is suboptimal on the erasure BC. This follows from the achievability in Corollary~\ref{cor:exact} and the  following Proposition~\ref{prop:loosened}. 

\begin{proposition}\label{prop:loosened}
For given $\M_1\geq 0$ and $\M:=\sum_{k=1}^K\M_k\geq 0$, 
	\begin{IEEEeqnarray}{rCl}\label{eq:loosened}
		\lefteqn{\C(\M_1, \ldots, \M_K) }  \nonumber \\
		& \leq&\min\bigg\{\C_{\K} + \frac{\M_1}{D}+\frac{(\M\!-\!\M_1)}{D}\cdot\!\frac{K \cdot \C_\K}{(K-1)\C_{\{2,\ldots, K\}}}, \nonumber \\
		 && \qquad \qquad \hspace{4.5cm} \C_1+ \frac{\M_1}{D} \bigg\} \IEEEeqnarraynumspace
	\end{IEEEeqnarray}
The RHS of \eqref{eq:loosened} is strictly less than $\C_\K+\frac{\M}{D}$ unless $\M=\M_1$ or $\delta_1=\ldots= \delta_K$.
\end{proposition}
\begin{IEEEproof}
See Appendix \ref{ap:smallMnotpossible}.
\end{IEEEproof}

\subsection{Noise-Free Bit-Pipe} \label{sec:sourcecoding}

Consider now the  noise-free bit-pipe model with uniform cache assignment in \cite{maddahali_niesen_2014-1}. It corresponds to an erasure BC where 
each receiver has zero erasure probability,
\begin{equation}\label{eq:perfect_channel}
\delta_1=\ldots=\delta_K=0.
\end{equation}
We adopt the ``source-coding perspective" of \cite{maddahali_niesen_2014-1}, and assume equal cache size 
\[
m_1=\cdots=m_K=m.
\]


From the upper bound  on $\C(\M_1, \ldots, \M_K)$ in Theorem~\ref{thm:upper_new}, the following \emph{lower bound} on the minimum achievable delivery rate $\rho^\star$ can be obtained as a function of the normalized symmetric cache size $m$:
\begin{corollary}\label{cor:minimumrate}
	For the noise-free bit-pipe model in \cite{maddahali_niesen_2014-1}:
\begin{IEEEeqnarray}{rCl}
\rho^\star \geq t-   m\cdot\min\bigg\{\frac{t^2 }{D},\;\sum_{k=1}^{t} \frac{k}{D-k+1} \bigg\}, \quad m\leq D. \IEEEeqnarraynumspace
\end{IEEEeqnarray}
\end{corollary}
\begin{IEEEproof} See Appendix~\ref{app:corBEC}.
\end{IEEEproof}
Figure~\ref{fig:source} compares this new converse result on $\rho$ with the existing converse results in \cite{maddahali_niesen_2014-1}, \cite{wanglimgastpar-2016},  and \cite{ghasemi_ramamoorthy}, and with the achievability result in \cite{yumaddahaliavestimehr-2016}. The converse result in \cite{ghasemi_ramamoorthy} is generally cumbersome to evaluate. The plot shows the numerical value calculated in \cite{ghasemi_ramamoorthy}.

\input{K12D64_sourcecoding}

\subsection{Gaussian BCs}\label{sec:Gaussian}
Finally, we specialize our results to memoryless Gaussian BCs. At time $t$, the
received symbol at receiver~$k$ is
\begin{equation}
Y_{k,t} = X_t + Z_{k,t}, 
\end{equation}
where $X_t$ is the input to the channel and $\{Z_{k,t}\}$ is an i.i.d. Gaussian process with zero mean and variance $\sigma_k^2>0$. The channel inputs are subject to an  average block-power constraint  $P$. The receivers are ordered in increasing strength:
\[
\sigma_1^2 \geq \sigma_2^2 \geq \ldots \geq \sigma_K^2>0.
\] 

By~\cite{bergmans74}, for every set $\set{S}$ as defined in \eqref{Aylin}, 
\begin{IEEEeqnarray}{rCl}\label{eq:common_SGaussian}
	\C_{ \set{S}}=\frac{1}{2}\log_2\left(1+\frac{\beta_1 P}{\sum_{k=2}^{|\set{S}|}\beta_{k}  P+\sigma_1^2}\right),
\end{IEEEeqnarray}
where $\beta_1,\ldots,\beta_{|\set{S}|}$ form the unique choice of $|\set{S}|$ real numbers in $[0,1]$ that  sum to $1$ and satisfy 
\begin{align}
\frac{\beta_1 P}{\sum_{k=2}^{|\set{S}|}\beta_k P+\sigma_1^2}= \frac{\beta_iP}{\sum_{k=i+1}^{{|\set{S}|}}\beta_k P+\sigma_{j_i}^2}, \quad i\in\{1,\ldots, |{S}|\}.
\end{align}
In particular, 
\begin{IEEEeqnarray}{rCl}
	\C_{k}= \frac{1}{2} \log\left(1+ \frac{P}{\sigma_k^2}\right), \quad k\in\{1,\ldots, K\}. 
\end{IEEEeqnarray}

Moreover, given a power constraint $P>0$, a  zero-mean variance-$P$ Gaussian input distribution $P_X$ maximizes $I(X;Y_k)$ and $I(X;Y_{k}|U)$ simultaneously for all $k\in \K$ and auxiliaries $U$ that form the Markov chain $U - X -Y_k$.  Therefore, Theorem~\ref{thm:upper_new}  and  Corollary~\ref{cor-equivalent}  coincide. Also,  
\begin{IEEEeqnarray}{rCl}
\C_{\textnormal{avg}} = \frac{1}{K} \sum_{k=1}^K 	\C_{k}= \frac{1}{K} \sum_{k=1}^K \frac{1}{2} \log\left( 1 +  \frac{P}{\sigma_k^2}\right).
\end{IEEEeqnarray}  

\begin{figure}[t!]
\input{K4D10Gaussian}
\caption{Upper and lower bounds on $\C^\star(\M)$ on a 4-receiver Gaussian BC with input power $P=1$ and noise variances $\sigma_1=4$, $\sigma_2^2=1$, $\sigma_3^2=0.5$, and $\sigma_4^2=0.1$. We have $D=10$.}
\label{fig:K4D10Gaussian}
\end{figure}
	Figure \ref{fig:K4D10Gaussian} shows the upper and lower bounds on $\C^{\star}(\M)$ in Propositions~\ref{thm:lower_global} and~\ref{thm:newglobal}.   The five blue points indicate the rate-memory points $(R^{(0)}, \M^{(0)})$, $(R^{\textsf{single}}, \M^{\textsf{single}})$, $(R^{(1)}, \M^{(1)})$, $(R^{(2)}, \M^{(2)})$, and $(R^{(3)}, \M^{(3)})$ for a zero-mean variance-$P$ Gaussian distribution $P_X$.  For comparison, the figure also shows the upper bound in Theorem~\ref{thm:upper_new} for a setup with uniform cache assignment $\frac{\M}{K}$ across all receivers. We observe that a smart cache assignment provides substantial gains in the capacity-memory tradeoff.

\section{Summary and Conclusion}\label{sec:summary}
We have provided close upper and lower bounds on the global capacity-memory tradeoff $\C^\star(\M)$ of degraded BCs. The bounds coincide in the regimes of small and large  total cache memory with thresholds depending on  the BC statistics. 
 For small cache memory sizes,  the weakest receiver needs to be assigned all. 
 In this regime, $\C^\star(\M)$ grows as $\frac{\M}{D}$, which corresponds to a perfect global caching gain where all receivers can benefit from  all the cache contents of the network. This performance is achieved by the proposed superposition piggyback coding scheme, which provides each receiver virtual access to the weakest receiver's cache contents. 
For the  regime of  moderate $\M$, we propose a generalized coded caching scheme, which assigns cache memories to all the receivers, with a larger cache memory the weaker a receiver is. Notice that  the larger the total cache budget $\M$, the larger the coded caching parameter $t\in\{1,\ldots, K-1\}$  needs to be chosen. This leads to a decreasing 
global caching gain because with increasing $t$ the various cache memories have more and more overlapping contents which cannot provide global caching gains. 
As a consequence, the slope of the rate-memory tradeoff achieved by generalized coded caching decreases with increasing total cache budget $\M$. The same  behaviour is also suggested   by the upper bound.  
For parameter $t=K-1$ generalized coded caching and the corresponding cache assignment exactly achieve the global capacity-memory tradeoff. Once  the total cache memory budget exceeds the corresponding cache budget, it is optimal to uniformly allocate all the remaining cache memory across all the receivers and to store the same content  in the extra portions of the receivers' cache memories. Here,  $\C^\star(\M)$ grows as $\frac{1}{K} \cdot \frac{\M}{D}$, which corresponds  to a local caching gain. 
We conclude that assigning the total cache memory uniformly across all the receivers is highly suboptimal over noisy BCs, in contrast to the noiseless setup considered in \cite{maddahali_niesen_2014-1}.

	\appendices
	
\section{Proof of Upper Bound in Theorem~\ref{thm:upper_new}} \label{sec:upperbound}

{Fix} the rate of communication 
\[
R < \C(\M_1,\ldots, \M_K).
\] 
Since $R$ is achievable, for each sufficiently large blocklength~$n$ and for each demand vector $\d$, there exist $K$  caching functions~$\big\{g_k^{(n)}\big\}$, an encoding function~$\{f_{\d}^{(n)}\}$, and $K$ decoding functions~$\big\{\varphi_{k,\d}^{(n)}\big\}$ so that the probability of worst-case error $\Pe^{(n)}(\mathbf{d})$ tends to 0 as $n\to \infty$. 

{Fix $\epsilon>0$ and a sufficiently large blocklength $n$ (depending on this $\epsilon$).  Let
\begin{IEEEeqnarray}{rCl}
\V_k&= &g_k^{(n)}(W_1, \ldots, W_D), \qquad k\in \{1,\ldots, K\}
  \end{IEEEeqnarray}
  denote the cache contents corresponding to the chosen caching function, and let for each demand vector $\d=(d_1,\ldots, d_K)$ with all different entries
\begin{IEEEeqnarray}{rCl}
X_\d^n&=&f_{\d}^{(n)}(W_1,\ldots, W_D) 
  \end{IEEEeqnarray}
  denote the input of the degraded BC corresponding to the chosen encoding functions.
Let $Y_{k, \d}^{n}$  denote the corresponding channel outputs at receiver~$k$.

\begin{lemma} \label{lem:upperbound}
There exist 
random variables $X_{\mathbf{d}},Y_{1, \mathbf{d}}, \ldots, Y_{K, \mathbf{d}}$ and for each set $\set{S}$ as in \eqref{Aylin} random variables $\{U_{\set{S}, 1,\d},\ldots, U_{\set{S},{|\set{S}|-1}, \d}\}$,  so that
{given $X_{\mathbf{d}}=x\in\set{X}$:}
\begin{subequations}\label{eq:conditions}
\begin{equation}\label{eq:auxiliary_channel}
(Y_{1, \mathbf{d}},  Y_{2, \mathbf{d}}, \ldots, Y_{K, \mathbf{d}})\sim \Gamma(\cdots|x);
\end{equation}
 and 
 for each $\set{S}$:
\begin{equation}\label{eq:MarkovU}
U_{\set{S}, 1,\d} - \cdots -  U_{\set{S},{|\set{S}|-1}, \d}- X_{\mathbf{d}} - Y_{K, \mathbf{d}} -  Y_{K-1, \mathbf{d}}\cdots  - Y_{1, \mathbf{d}}
\end{equation}
  \end{subequations}
forms a Markov chain 
and the following $|\set{S}|$ inequalities hold:
\begin{subequations}\label{eq:d_inequa}
\begin{align}
R-\epsilon&\leq \frac{1}{n} I\big(W_{d_{j_1}};\V_{j_1}\big) +I\big(U_{\set{S},1,\d};Y_{j_1,\mathbf{d}} \big),\\
R-\epsilon&\leq\frac{1}{n} I\big(W_{d_{j_k}};\V_{j_1}, \ldots, \V_{j_k}| W_{d_{j_1}},\ldots, W_{d_{j_{k-1}}}\big)\nonumber \\& \qquad +I\big(U_{\set{S},k,\d};Y_{j_k,\mathbf{d}} | U_{\set{S},k-1,\d}), \,\, 
\nonumber \\ & \hspace{3.5cm}\forall k\in\{2,\ldots, |\set{S}|-1\}, \\ 
R-\epsilon&\leq\frac{1}{n} I\big(W_{d_{j_{|\set{S}|}}};\V_{j_1}, \ldots, \V_{j_{|\set{S}|}}| W_{d_{j_1}},\ldots, W_{d_{j_{|\set{S}|-1}}}\big) \nonumber \\
 & \qquad +I\big(X_{\d};Y_{j_{|\set{S}|},\mathbf{d}} | U_{\set{S},|\set{S}|-1,\d}).
\end{align}
\end{subequations}
\end{lemma}
\begin{IEEEproof}
The proof is similar to the converse proof of the capacity of degraded BCs without caching \cite{gallager74}. 

Since the worst case error probability is bounded by $\epsilon$, using Fano's inequality we have
\begin{subequations}\label{eq:out1}
\begin{align}\label{eq:36a}
R\!-\!\epsilon &\leq\frac{1}{n} I\big(W_{d_{j_1}};Y_{j_1,\mathbf{d}}^n,\V_{j_1}\big) \nonumber\\
& =  \frac{1}{n} I\big(W_{d_{j_1}};\V_{j_1}\big)+\frac{1}{n} I\big(W_{d_{j_1}};Y_{j_1,\mathbf{d}}^n\big|\V_{j_1}\big).
\end{align}
Similarly, for $k\in\{2,\ldots, K\}$:
\begin{align}\label{eq:36b}
R\!-\!\epsilon \stackrel{(a)}{\leq}&\frac{1}{n} I\big(W_{d_{j_k}};Y_{{j_k},\mathbf{d}}^n,\V_{{j_1}}, \ldots, \V_{j_k}\big| W_{d_{{j_1}}},\ldots, W_{d_{{j_{k-1}}}}\big)\nonumber \\
 =& \frac{1}{n} I\big( W_{d_{{j_k}}}; \V_{j_1}, \ldots, \V_{j_k} \big| W_{d_1},\ldots, W_{d_{{j_{k-1}}}}\big) \nonumber\\&+ \frac{1}{n} I\big(W_{d_{j_k}};Y_{{j_k},\mathbf{d}}^n \big| \V_{1}, \ldots, \V_{j_k}, W_{d_{j_1}},\ldots, W_{d_{{j_{k-1}}}}\big),
\end{align}
\end{subequations}
where $(a)$ uses Fano's inequality as well as the fact that all messages are independent. Recall that the demand vector $\d$ has all different entries.

We next develop the second summands in \eqref{eq:36a} and \eqref{eq:36b}. For  the second summand in \eqref{eq:36a} we write
\begin{align}
\frac{1}{n} I\big(W_{d_{j_1}};Y_{j_1,\mathbf{d}}^n\big|\V_{j_1}\big)
& =   \frac{1}{n}\sum_{t=1}^n  I\big(W_{d_{j_1}};Y_{j_1,\mathbf{d},t}\big|\V_{j_1},Y_{j_1,\mathbf{d}}^{t-1} \big)\nonumber \\
 & \leq    \frac{1}{n}\sum_{t=1}^n I\big(W_{d_{j_1}}, Y_{j_1,\mathbf{d}}^{t-1};Y_{j_1,\mathbf{d},t}\big|\V_{j_1} \big) \nonumber \\
 &  =   I\big({U}_{\set{S}, 1,\d,T};Y_{j_1,\mathbf{d},T}\big|\V_{j_1},T\big) \nonumber \\
&    \leq   I\big({U}_{\set{S},1,\d};Y_{j_1,\mathbf{d}}\big|\V_{j_1}\big)\label{eq:out0}
\end{align}
where $T$ denotes a random variable that is uniformly distributed over $\{1,\ldots, n\}$ and independent of all previously defined random variables, and where 
\begin{IEEEeqnarray*}{rCl}
U_{\set{S},1,\d,T}&:=&(\V_{j_1}, W_{d_{j_1}}, Y_{{j_1}, \mathbf{d}}^{t-1}), \\
U_{\set{S},1,\d} & := & (U_{\set{S},{1},\d,T}, T), \\
Y_{j_1,\mathbf{d}} &:=&Y_{{j_1},\mathbf{d},T}.
\end{IEEEeqnarray*}
Define further for $k\in\{2,\ldots, |\set{S}|-1\}$:
\begin{IEEEeqnarray*}{rCl}
U_{\set{S},k,\d,T} & := & (U_{\set{S}, k-1, \d, T}, \V_{j_k}, W_{d_{j_k}}, Y_{j_k, \mathbf{d}}^{t-1}), \\
U_{\set{S},k,\d}& := & U_{\set{S},k,\d,T},\\
Y_{j_k,\mathbf{d}} &:=&Y_{j_k,\mathbf{d},T}, 
\end{IEEEeqnarray*}
and 
\begin{IEEEeqnarray*}{rCl}
Y_{j_{|\set{S}|},\mathbf{d}}& := &Y_{j_{|\set{S}|},\mathbf{d},T}\\
X_{\mathbf{d}}& := &X_{\mathbf{d},T}.
\end{IEEEeqnarray*}
For $k\in\{2,\ldots, K-1\}$, we  expand the second summand in \eqref{eq:36b} as:
\begin{align}
&{\frac{1}{n} I\big(W_{d_{j_k}};Y_{{j_k},\mathbf{d}}^n \big| \V_{{j_1}}, \ldots, \V_{j_k}, W_{d_{j_1}},\ldots, W_{d_{{j_{k-1}}}}\big) } \qquad \nonumber \\
 &\, =  \frac{1}{n} \sum_{t=1}^n I\big(W_{d_{j_k}};Y_{{j_k},\mathbf{d},t} \big| \V_{{j_1}}, \ldots, \V_{j_k}, W_{d_{j_1}},\ldots, W_{d_{{j_{k-1}}}}, Y_{{j_k},\mathbf{d}}^{t-1}\big) \nonumber \\
&\,\stackrel{(a)}{=}  \frac{1}{n} \sum_{t=1}^n I\big(W_{d_{j_k}};Y_{{j_k},\mathbf{d},t} \big| \V_{{j_1}}, \ldots, \V_{j_k}, W_{d_{j_1}},\ldots, W_{d_{{j_{k-1}}}}, \nonumber \\
& \hspace{5.2cm} Y_{{j_1},\mathbf{d}}^{t-1}, \ldots, Y_{{j_{k-1}},\mathbf{d}}^{t-1}, Y_{{j_k},\mathbf{d}}^{t-1}\big) \nonumber \\
&\,\leq \frac{1}{n}\sum_{t=1}^n I\big( W_{d_{j_k}},Y_{{j_k},\mathbf{d}}^{t-1} , \V_{j_k}; Y_{{j_k},\mathbf{d},t} \big| \V_{{j_1}}, \ldots, \V_{j_{k-1}}, \nonumber \\
&  \hspace{3.2cm} 
W_{d_{j_1}},\ldots, W_{d_{{j_{k-1}}}},Y_{{j_1},\mathbf{d}}^{t-1}, \ldots, Y_{{j_{k-1}},\mathbf{d}}^{t-1}\big)\nonumber\\
&\,={ I\big(U_{\set{S},{k},\d,T};Y_{{j_k},\mathbf{d},T} \big| U_{\set{S},{{k-1}},\d,T}, T)}\nonumber \\
&\,= {I\big(U_{\set{S},{k},\d};Y_{{j_k},\mathbf{d}} \big| U_{\set{S},{{k-1}},\d})},\label{eq:out2}
\end{align}
where (a) follows from the degradedness of the outputs.

Similarly, we also have 
\begin{align}
&{\frac{1}{n} I\big(W_{d_{j_{|\set{S}|}}};Y_{{j_{|\set{S}|}},\mathbf{d}}^n \big| \V_{{j_1}}, \ldots, \V_{j_{|\set{S}|}}, W_{d_{j_1}},\ldots, W_{d_{{j_{{|\set{S}|}-1}}}}\big) } \qquad \nonumber \\
&\,= \frac{1}{n} \sum_{t=1}^n I\big(W_{d_{j_{|\set{S}|}}};Y_{{j_{|\set{S}|}},\mathbf{d},t} \big| \V_{{j_1}}, \ldots, \V_{j_{|\set{S}|}}, W_{d_{j_1}},\ldots, W_{d_{{j_{{|\set{S}|}-1}}}}, \nonumber \\
& \hspace{4.8cm} Y_{{j_1},\mathbf{d}}^{t-1}, \ldots, Y_{{j_{{|\set{S}|}-1}},\mathbf{d}}^{t-1}, Y_{{j_{|\set{S}|}},\mathbf{d}}^{t-1}\big) \nonumber \\
&\,\leq \frac{1}{n}\sum_{t=1}^n I\big( W_{d_{j_{|\set{S}|}}},Y_{{j_{|\set{S}|}},\mathbf{d}}^{t-1} , \V_{j_{|\set{S}|}}; Y_{{j_{|\set{S}|}},\mathbf{d},t} \big| \V_{{j_1}}, \ldots, \V_{j_{{|\set{S}|}-1}}, \nonumber \\
&  \hspace{3cm} 
W_{d_{j_1}},\ldots, W_{d_{{j_{{|\set{S}|}-1}}}},Y_{{j_1},\mathbf{d}}^{t-1}, \ldots, Y_{{j_{{|\set{S}|}-1}},\mathbf{d}}^{t-1}\big)\nonumber\\
&\,\leq{ I(X_{\d,T};Y_{{j_{|\set{S}|}},\mathbf{d},T}\  | \ U_{\set{S},{{{|\set{S}|}-1}},\d,T}, T)}\nonumber \\
&\,= {I(X_{\d};Y_{{j_{|\set{S}|}},\mathbf{d}}\  | \ U_{\set{S},{{{|\set{S}|}-1}},\d})}.\label{eq:out4}
\end{align}

It can be verified that the defined random variables satisfy Conditions~\eqref{eq:conditions}. Combining this observation with \eqref{eq:out1}--\eqref{eq:out4} concludes the proof.
\end{IEEEproof}
\vspace{2mm}

We average the bounds in \eqref{eq:d_inequa} over demand vectors. Let $\set{Q}^{\textnormal{dist}}_K$ be the set of all the  ${D \choose K}{K!}$ $K$-dimensional demand vectors with all distinct entries. Also, let $Q$ be a uniform random variable over the elements of $\set{Q}^{\textnormal{dist}}_K$ and independent of all other random variables. 
Define for each set $\set{S}$ as in \eqref{Aylin}: $U_{\set{S},1}:=(U_{\set{S},{1}, Q}, Q)$;  $U_{\set{S},k}:=U_{\set{S},k,Q}$, for $k\in\{2,\ldots, |\set{S}|-1\}$; $X:=X_{Q}$; and $Y_k:=Y_{k,Q}$ for $k\in\K$.

  Notice that the defined random variables satisfy conditions~\eqref{eq:markov_channel} and 	\eqref{eq:markovsetS} in the theorem. It remains to prove that they also satisfy~\eqref{eq:d_inequa10}. To this end, we average
inequalities~\eqref{eq:d_inequa} over all the demand vectors in~$\set{Q}^{\textnormal{dist}}_K$. Using standard {arguments to take care of the time-sharing random variable~$Q$}, 
and defining 
\begin{subequations}\label{defgen}
\begin{align}
{\alpha}_{\set{S},1}&\!:=\! \frac{1}{{D \choose K}{{K!}}}\sum_{\d \in \set{Q}^{\textnormal{dist}}_K }\!\frac{1}{n} I(W_{d_{j_1}};\V_{1}), \\
{\alpha}_{\set{S},k}&\!:=\! \frac{1}{{D \choose K}{{K!}}}\sum_{\d \in \set{Q}^{\textnormal{dist}}_K}\!\frac{1}{n} I(W_{d_{j_k}};\V_{1}, \ldots, \V_{j_k}|W_{d_{j_1}}, \ldots, W_{d_{j_{k\hspace{-.05cm}-\hspace{-.05cm}1}}}\hspace{-.05cm}), \nonumber \\
& \hspace{4.3cm} k\in\{2,\ldots, |\set{S}|\},
\end{align}
\end{subequations}
we obtain for each $\set{S}$ as in \eqref{Aylin}:
\begin{subequations}\label{eq:d_inequa2}
\begin{align}
R-\epsilon\leq&\ I\big(U_{\set{S},1};Y_{j_1} \big) + \alpha_{\set{S},1},\\
R-\epsilon\leq&\ I\big(U_{\set{S},k};Y_{j_k} | U_{\set{S},k-1})+  \alpha_{\set{S},k}, \quad \forall k\!\in\!\{2,\ldots, |\set{S}|-1\}, \\
R-\epsilon\leq& \ I\big(X;Y_{j_{|\set{S}|}} | U_{\set{S},|\set{S}|-1})+ \alpha_{\set{S},|\set{S}|},
\end{align}
\end{subequations}

\begin{lemma}\label{lem:alphas}
For each set $\set{S}$, parameters $\alpha_{\set{S},1}, \ldots, \alpha_{\set{S},|\set{S}|}$ satisfy  the following constraints:
\begin{subequations} \label{eq:cons}
\begin{align}
0 \leq {\alpha}_{{\set{S},k}} &\leq  \frac{{\sum_{i=1}^k \M_{j_i}}}{D-k+1}, \qquad k\in\{1,\ldots, |\set{S}|\},\label{eq:cons1} \\
{\alpha}_{\set{S},k'} &\leq{\alpha}_{{\set{S},k}}, \qquad  k, k' \in\{1,\ldots, |\set{S}|\}, \ k' \leq k,\label{eq:cons2}\\
\label{eq:cons3prime}
\sum_{k=1}^{|\set{S}|} {\alpha}_{{\set{S},k}} &  \leq \frac{|\set{S}|}{D}\sum_{k=1}^{|\set{S}|} \M_{j_k}.
\end{align} 
\end{subequations}

\end{lemma}
\begin{IEEEproof}
See Appendix \ref{app:lemalphas}.
\end{IEEEproof}

By \eqref{eq:d_inequa2}--\eqref{eq:cons} and letting $\epsilon \to 0$, the following intermediate result---which is used in other proofs in this paper---is obtained.
\begin{lemma} \label{lem:intermediate}
There exist random variables $X, Y_1, \ldots, Y_K$ and  for \emph{every} receiver set $\set{S}$ as in \eqref{Aylin} random variables $\{U_{\set{S},1}, \ldots, U_{\set{S},{|\set{S}|-1}} \}$, so that  \eqref{eq:markov_channel} and 	\eqref{eq:markovsetS} hold, and for all $\set{S}$:
\begin{subequations}\label{eq:d_inequa5}
	\begin{align}
	\C(\M_1,\ldots, \M_K)\leq&I\big(U_{\set{S},1};Y_{j_1} \big)+ \alpha_{{\set{S},1}} ,\\
	\C(\M_1,\ldots, \M_K)\leq&I\big(U_{\set{S},k};Y_{j_k} | U_{\set{S},k-1})+ \alpha_{{\set{S},k}},  \nonumber \\
	 & \hspace{2cm} \forall k\in\{2,\ldots, |\set{S}|\}, 
	\end{align}
	for parameters  $\alpha_{{\set{S},1}}, \ldots, \alpha_{{\set{S},|\set{S}|}}$ satisfying~\eqref{eq:cons}.
	\end{subequations}
\end{lemma}

By the following Lemma~\ref{lem:sym_alphas},  because constraints~\eqref{eq:d_inequa5} are increasing in $\alpha_{\set{S},1},\ldots, \alpha_{\set{S},|\set{S}|}$,  and by constraint~\eqref{eq:cons3prime}, we conclude  that the choice $\alpha_{\set{S},k}= \alpha_{\set{S},k}^\star$ in \eqref{eq:values_of_alpha} makes the upper bound \eqref{eq:d_inequa5} loosest. The following Lemma~\ref{lem:sym_alphas} thus concludes the proof.

\begin{lemma}\label{lem:sym_alphas}
Lemma~\ref{lem:intermediate} remains valid, if parameters $\alpha_{\set{S},1}, \ldots, \alpha_{\set{S},|\set{S}|}$ are further constrained to satisfy for each $k\in\{1,\ldots, |\set{S}|-1\}$  one of the two following conditions: 
\begin{itemize}
	\item $\alpha_{\set{S},k}= \frac{{\sum_{i=1}^k \M_{j_i}}}{D-k+1}$; or
	\item $\alpha_{\set{S},k}=\alpha_{\set{S},k+1}$. 
\end{itemize}
\end{lemma}
\begin{IEEEproof}
	See Appendix~\ref{app:lemma_sym_alphas}.
\end{IEEEproof}

\section{Proof of Lemma~\ref{lem:alphas}}\label{app:lemalphas}
We only prove the lemma for $\set{S}=\K$. The other proofs are similar. 

We first prove~\eqref{eq:cons1}. Every
$\alpha_{\set{K},k}$ is non-negative, because mutual information is non-negative. To prove the upper bound in \eqref{eq:cons1}, we proceed as follows.	Let $\mathcal{Q}_K^{\textnormal{dist}}$ be the set of  $K$-dimensional demand vectors that have $K$ distinct entries in $\{1,\ldots, D\}$; and for each $k\in\{1,\ldots, K\}$ and each $k-1$ dimensional demand vector $\tilde{\d}=(d_1,\ldots, d_{k-1})$, define  $W_{\tilde{\d}}:=(W_{d_1}, \ldots, W_{d_{k-1}})$. We have:
	\begin{align}
	&\alpha_{\K,k}\nonumber \\
	&=\frac{1}{K!{D \choose K}}\sum_{\mathbf{d}\in\set{Q}_K^{\textnormal{dist}}}I(W_{d_k};\V_1,\ldots, \V_k|W_{d_1},\ldots,W_{d_{k-1}})\nonumber\\
	&=\frac{1}{K!{D \choose K}} \sum_{\tilde{\d}\in \Qkone} \sum_{\substack{\mathbf{d}\in\set{Q}_K^{\textnormal{dist}} \colon \\ (d_1,\ldots, d_{k-1})=\tilde{\d}}} I(W_{d_k};\V_1,\ldots, \V_k|W_{\tilde{\d}})\nonumber\\
	 &\stackrel{(a)}{=}\frac{1}{K!{D \choose K}} \sum_{\tilde{\d}\in \Qkone} \; \sum_{j \in\set{D}\backslash \tilde{\d}} I(W_{j};\V_1,\ldots, \V_k|W_{\tilde{\d}}) \nonumber \\
	  &\hspace{3.4cm} \cdot {{D-k} \choose {K-k}} (K-k)!\nonumber\\
			&=\frac{1}{{k!}{D\choose k}} \sum_{\tilde{\d}\in \Qkone} \; \sum_{j \in\set{D}\backslash \tilde{\d}}  I(W_{j};\V_1,\ldots, \V_k |W_{\tilde{\d}})\nonumber\\
	&\stackrel{(b)}{=}\frac{1}{{k!}{D\choose k}} \sum_{\tilde{\d}\in \Qkone} \big[ H(W_1,\ldots, W_N|W_{\tilde{\d}}) \nonumber \\
	 & \hspace{3cm} -\sum_{j \in\set{D}\backslash \tilde{\d}}  H(W_{j}|\V_1,\ldots, \V_k , W_{\tilde{\d}}) \big]\nonumber\\
	&\stackrel{(c)}{\leq}\frac{1}{{k!}{D \choose k}}\sum_{\tilde{\d}\in \Qkone} I(W_1,\ldots, W_N;\V_1,\ldots, \V_k|W _{\tilde{\d}}) \nonumber\\
	&\stackrel{(d)}{\leq} \frac{(k-1)!{D \choose k-1}}{{k!}{D \choose k}}\sum_{i=1}^k \M_i\nonumber\\
	&={\frac{ \sum_{i=1}^k \M_i}{D-k+1}}
	\end{align}
	where $(a)$ holds because for each value of $K$ and $j$ there are ${{D-k} \choose {K-k}} (K-k)!$ ordered demand vectors $\d\ \in \set{Q}_K^{\textnormal{dist}}$ with $(d_1,\ldots, d_{k-1})=\tilde{\d}$ and with $d_k=j$;  (b) holds  by the independence of the messages; (c) holds because for any random tuple $(A_1, \ldots, A_L)$ it holds that $\sum_{l=1}^L H(A_l) \geq H(A_1, \ldots, A_L)$; and (d) holds because $ I(W_1,\ldots, W_N;\V_1,\ldots, \V_k|W _{\tilde{\d}})$ cannot exceed {$\sum_{i=1}^k \M_i$}. This concludes the proof of  \eqref{eq:cons1}.

\renewcommand{\overrightarrow}[1]{#1}

To prove constraint~\eqref{eq:cons2}, we fix a $K$-dimensional demand vector~$\d\in \set{Q}_K^{\textnormal{dist}}$, and consider the cyclic shifts of this vector. For $\ell\in\{0,\ldots, K-1\}$, let $\overrightarrow{\d}^{(\ell)}$ be the vector obtained from $\overrightarrow{\d}$ when the elements are cyclically shifted $\ell$ positions to the right. (For example, if $\d=(1, 2, 3)$ then $\overrightarrow{\d}^{(2)}=(2, 3, 1)$.) 
For each $\ell\in\{0,\ldots, K-1\}$ and $k\in\{1,\ldots, K\}$, let $\overrightarrow{d}_{k}^{(\ell)}$ denote the $k$-th index of demand vector $\overrightarrow{\d}^{(\ell)}$. So, 
\begin{equation}\label{eq:circ}
\overrightarrow{d}_{k}^{(\ell)} = d_{(k-\ell)\!\! \!\!\mod K}
\end{equation}
where for each positive integer $\xi$ the term $(\xi \mod K)$ takes value in $\{1,\ldots, K\}$ so that 
{\begin{equation}
\xi \mod K =
\xi - b K \quad  \textnormal{ for some positive integer } b.
\end{equation}}

For each $\ell\in\{1,\ldots, K\!-\!1\}$ and $k, k'\in\{2,\ldots, K\}$ with $k'\leq k$, we write
\begin{align}\label{eq:in1}
\lefteqn{I(W_{d_1};\V_{1}) {\stackrel{(a)}{=}  I(W_{\overrightarrow{d}_{k'}^{(k'-1)}};\V_{1})}}\quad\nonumber\\
 \stackrel{(b)}{\leq}  &  I(W_{\overrightarrow{d}_{k'}^{(k'-1)}};\V_{1}\!\!, \ldots,\V_{k'}| W_{\overrightarrow{d}_1^{(k'-1)}},\ldots, W_{\overrightarrow{d}_{k'-1}^{(k'-1)}}) \nonumber \\
  \stackrel{(a)}{=}  &{  I(W_{\overrightarrow{d}_{k}^{(k-1)}};\V_{1}\!\!, \ldots, \V_{k'}| W_{\overrightarrow{d}_{1+ k-k'}^{(k-1)}},\ldots, W_{\overrightarrow{d}_{k-1}^{(k-1)}}) }\nonumber \\
  \stackrel{(b)}{\leq} &  I(W_{\overrightarrow{d}_{k}^{(k-1)}};\V_{1}\!\!, \ldots, \V_k| W_{\overrightarrow{d}_1^{(k-1)}},\ldots, W_{\overrightarrow{d}_{k-1}^{(k-1)}}) 
\end{align}
{where (a) follows by \eqref{eq:circ} and (b) is by the independence of  messages.}  

{Fix a demand vector $\d\in\set{Q}_K^{\textnormal{dist}}$ and sum up the above  inequality~\eqref{eq:in1} over all $K$ cyclic shifts $\d^{(0)}, \d^{(1)}, \ldots,$ $\d^{(K-1)}$ of $\d$ to obtain:}
\begin{IEEEeqnarray}{rCl}
\lefteqn{\sum_{\ell=0}^{K-1} I(W_{\overrightarrow{d}_1^{(\ell)}};\V_{1})}\quad\nonumber\\& \leq &  \sum_{\ell=0}^{K-1} I(W_{\overrightarrow{d}_{k'}^{(\ell)}};\V_{1}, \ldots, \V_{k'}| W_{\overrightarrow{d}_1^{(\ell)}},\ldots, W_{\overrightarrow{d}_{k'-1}^{(\ell)}}) \nonumber \\
 & \leq & \sum_{\ell=0}^{K-1}  I(W_{\overrightarrow{d}_{k}^{(\ell)}};\V_1, \ldots, \V_k| W_{\overrightarrow{d}_1^{(\ell)}},\ldots, W_{\overrightarrow{d}_{k-1}^{(\ell)}}). \label{eq:in2}\IEEEeqnarraynumspace
 \end{IEEEeqnarray}
Since the set $\set{Q}_K^{\textnormal{dist}}$ can be partitioned into subsets of demand vectors that are cyclic shifts of each others and all cyclic shifts of a demand vector in $\set{Q}_K^{\textnormal{dist}}$ are also in $\set{Q}_K^{\textnormal{dist}}$,   {we  conclude from~\eqref{eq:in2}:}
\begin{IEEEeqnarray}{rCl}
\lefteqn{\sum_{\d\in \set{Q}_K^{\textnormal{dist}}} I(W_{d_1};\V_{1})}\quad\nonumber\\& \leq &  \sum_{\d\in\set{Q}_K^{\textnormal{dist}}} I(W_{d_{k'}};\V_{1}, \ldots, \V_{k'}| W_{d_1},\ldots, W_{d_{k'-1}}) \nonumber \\
 & \leq & \sum_{\d\in\set{Q}_K^{\textnormal{dist}}} I(W_{d_{k}};\V_{1}, \ldots, \V_k| W_{d_1},\ldots, W_{d_{k-1}}).
 \end{IEEEeqnarray}
{This proves~\eqref{eq:cons2}.}

 We proceed to prove constraint~\eqref{eq:cons3prime}. 
 For each  $\d\in \set{Q}_K^{\textnormal{dist}}$: 
\begin{IEEEeqnarray}{rCl}
\lefteqn{I(W_{d_1}; \V_1)+ \sum_{k=2}^K I(W_{d_k}; \V_1, \ldots, \V_k|W_{d_1}, W_{d_2}, \ldots, W_{d_{k-1}})} \quad \nonumber \\
&  \leq & I(W_{d_1},W_{d_2}, \ldots, W_{d_{K}}; \V_1, \ldots, \V_K). \hspace{2.3cm}
\end{IEEEeqnarray}
So,
\begin{align}
&\sum_{\d\in\set{Q}_K^{\textnormal{dist}}} \bigg[ I(W_{d_1}; \V_1) \nonumber \\
& \qquad \qquad +\sum_{k=2}^K I(W_{d_{k}}; \V_1, \ldots, \V_k|W_{d_1}, W_{d_2}, \ldots, W_{d_{k-1}}) \bigg]  \qquad \nonumber \\
 &\quad  \leq   \sum_{\d \in \set{Q}_K^{\textnormal{dist}}} I(W_{d_1}, W_{d_2},\ldots, W_{d_{K}}; \V_1 \ldots, \V_K) \nonumber \\
 &\quad \stackrel{(a)}{=}  \sum_{\d \in \set{Q}_K^{\textnormal{dist}}}  \Big[ H(W_{d_1})+ H(W_{d_2})+ \ldots + H(W_{d_K})  \nonumber\\&\qquad\qquad\qquad- H( W_{d_1}, \ldots, W_{d_K}| \V_1, \ldots, \V_K) \Big]\nonumber \\
&\quad  \stackrel{(b)}{=}  \frac{K}{D} |\set{Q}_K^{\textnormal{dist}}| H(W_1, \ldots, W_D)\nonumber\\&\qquad\qquad -  \sum_{\d \in\set{Q}_K^{\textnormal{dist}}}  H( W_{d_1}, \ldots, W_{d_K}| \V_1, \ldots, \V_K)  \nonumber \\
 &\quad \stackrel{(c)}{\leq}     \frac{K}{D}{K!} {D \choose K}  H(W_1, \ldots, W_D)  \nonumber\\&\qquad \qquad-   \frac{K}{D}{K!}  {D \choose K} H( W_1, \ldots, W_D| \V_1, \ldots, \V_K) \nonumber \\
&\quad  \stackrel{(b)}{=}   \frac{K}{D}{K!} {D \choose K} I( W_1, \ldots, W_D; \V_1, \ldots, \V_K)\nonumber\\
&\quad \leq    \frac{K}{D}{K!} {D \choose K}{n}\sum_{k=1}^K M_k, \nonumber
\end{align}
where (a) holds by the chain rule of mutual information,  (b) by the independence and uniform rate of messages $W_1,\ldots, W_D$ and the definition of the set $\set{Q}^{\textnormal{dist}}_K$, which is of size {${D \choose K} K!$}, and (c) by the generalized Han-Inequality (the following Proposition~\ref{prop:han}).
\vspace{1mm}

\begin{proposition} \label{prop:han}
Let $L$ be a positive integer and $A_1,\ldots, A_L$ be a finite random $L$-tuple. Denote by  $A_{\mathcal{J}}$ the subset $\{A_l,\ l\in\mathcal{J}\}$.
For every $i \in\{1,\ldots, L\}$:
\begin{align}
{1\over {L \choose i}}\sum_{\substack{\mathcal{J}\subseteq\{1,\ldots,L\}:\\|\mathcal{J}|=i}}\frac{H(A_{\mathcal{J}})}{i}\geq \frac{1}{L}H(A_1,\ldots,A_L).\label{Han}
\end{align}
\end{proposition}
\begin{IEEEproof}
See~\cite[Theorem 17.6.1]{CoverThomas}.
\end{IEEEproof}

	\section{Proof of Lemma~\ref{lem:sym_alphas}}\label{app:lemma_sym_alphas}

We prove the lemma by contradiction. Fix a random tuple $(X, Y_1, \ldots, Y_K)$ satisfying \eqref{eq:markov_channel} and for each set $\set{S}$ as in \eqref{Aylin} a random tuple  $U_{\set{S},1}, U_{\set{S},2}, \ldots, U_{\set{S},|\set{S}-1}$ satisfying  \eqref{eq:markovsetS}  and real numbers ${\alpha}_{\set{S},1}, \ldots,{\alpha}_{\set{S},|\set{S}|}$  satisfying \eqref{eq:cons}. 

Assume that for some set $\set{S}$ as in \eqref{Aylin} and some $\tilde{k}\in\{1,\ldots, |\set{S}|-1\}$: 
\begin{equation}
\alpha_{\set{S},\tilde{k}} \neq \alpha_{\set{S},\tilde{k}+1}
\end{equation} 
and 
\begin{equation}\label{eq:co2}
	 \alpha_{\set{S},\tilde{k}} < \frac{\sum_{i=1}^{\tilde{k}} \M_{j_i}}{D-\tilde{k}+1}.
\end{equation}
Let  
  \begin{equation}
  \gamma:=\max\Bigg\{ \frac{1}{2},\ \frac{ \alpha_{\set{S},\tilde{k}+1}-\frac{\sum_{i=1}^{\tilde{k}} \M_{j_i}}{D-\tilde{k}+1} }{\alpha_{\set{S},\tilde{k}+1}- \alpha_{\set{S},\tilde{k}}}\Bigg\}.
  \end{equation}
  Notice that by \eqref{eq:co2}:
  \begin{equation}
  \gamma \in \left[\frac{1}{2},\, 1\right).
  \end{equation}
Define 
  the new parameters
  \begin{subequations}\label{eq:baralpha}
  	\begin{IEEEeqnarray}{rCl}
  		\bar{\alpha}_{\set{S},k}&: = &\alpha_{\set{S},k}, \quad k\in\{1,\ldots, |\set{S}|\}\backslash \{\tilde{k}, \tilde{k}+1\}\\
  		\bar{\alpha}_{\set{S},\tilde{k}} &:= &\gamma \alpha_{\set{S},\tilde{k}}+(1-\gamma)\label{eq:bark} \alpha_{\set{S},\tilde{k}+1}\\
  		\bar{\alpha}_{\set{S},\tilde{k}+1} &:= &(1-\gamma )\alpha_{\set{S},\tilde{k}}+\gamma \alpha_{\set{S},\tilde{k}+1}.\label{eq:barkplus}
  	\end{IEEEeqnarray} 
  \end{subequations}
Notice that this new set of parameters  satisfies constraints \eqref{eq:cons} when $\alpha_{\set{S},1}, \ldots, \alpha_{\set{S},|\set{S}|}$ are replaced by $\bar{\alpha}_{\set{S},1}, \ldots, \bar{\alpha}_{\set{S},|\set{S}|}$. In particular, 
\begin{equation}\label{eq:order}
\bar{\alpha}_{\set{S},{k}} \leq  \bar{\alpha}_{\set{S},{k}+1}, \qquad k\in\{1,\ldots, |\set{S}|-1\}.
\end{equation}
 
  We will show that there exist  new auxiliary random variables $\bar{U}_{\set{S},1}, \bar{U}_{\set{S},2}, \ldots, \bar{U}_{\set{S},|\set{S}|-1}$ 
  satisfying the Markov chain \eqref{eq:markovsetS}, and so that 
  upper bound \eqref{eq:d_inequa}  is looser for  these new auxiliares and the new parameters $\bar{\alpha}_{\set{S},1},\ldots, \bar{\alpha}_{\set{S},|\set{S}|}$
 than  for the original auxiliaries $U_{\set{S},1}, \ldots, U_{\set{S}, |\set{S}|-1}$ and parameters $\alpha_{\set{S},1},\ldots, \alpha_{\set{S},|\set{S}|-1}$.

To simplify notation in the following, we define
\begin{equation}
U_{\set{S},|\set{S}|}:=X.
\end{equation}
Notice that since $\alpha_{\set{S},\tilde{k}} \neq \alpha_{\set{S},\tilde{k}+1}$ and by \eqref{eq:cons2}, the strict inequality 
\begin{equation}\label{eq:strict}
\alpha_{\set{S},\tilde{k}} < \alpha_{\set{S},\tilde{k}+1}
\end{equation} must hold.
Choose
\begin{IEEEeqnarray}{rCl}
\bar{U}_{\set{S},k} &= & U_{{S},k}, \qquad  k \in\{1,\ldots,|\set{S}|-1\} \backslash \{\tilde{k}\},
\end{IEEEeqnarray}
and 
\begin{equation}
\bar{U}_{\set{S},|\set{S}|} =  U_{\set{S},|\set{S}|}=X.
\end{equation}
The choice of $\bar{U}_{\set{S},\tilde{k}}$ depends on whether 
\begin{subequations}
\begin{equation}\label{eq:condition1}
I(U_{\set{S},\tilde{k}};Y_{\tilde{k}}| U_{\set{S},\tilde{k}-1}) \leq I(U_{\set{S},\tilde{k}+1};Y_{\tilde{k}+1}|U_{\set{S},\tilde{k}}),
\end{equation} 
or 
\begin{equation}\label{eq:condition2}
I(U_{\set{S},\tilde{k}};Y_{\tilde{k}}|U_{\set{S},\tilde{k}-1}) > I(U_{\set{S},\tilde{k}+1};Y_{\tilde{k}+1}|U_{\set{S},\tilde{k}}).
\end{equation}
\end{subequations}
If \eqref{eq:condition1} holds, choose
\begin{equation}
\bar{U}_{\set{S},\tilde{k}} =  U_{\set{S},\tilde{k}}.
\end{equation}
If \eqref{eq:condition2} holds, let $E \in \{0,1\}$ be a Bernoulli-$\beta$ random variable independent of everything else, where
 \begin{equation}\label{eq:beta}
\beta :=(1-\gamma)- (1-\gamma) \cdot\frac{I(U_{\set{S},\tilde{k}+1};Y_{\tilde{k}+1}|U_{\set{S},\tilde{k}})}{ I(U_{\set{S},\tilde{k}};Y_{\tilde{k}}|U_{\set{S},\tilde{k}-1})}.
 \end{equation}
 Choose
 \begin{equation}
 \bar{U}_{\set{S},\tilde{k}} = \begin{cases} (U_{\set{S},{\tilde{k}}}, E),& \ \textnormal{if } E=0\\
(U_{\set{S},\tilde{k}-1},E), & \ \textnormal{if } E=1.\end{cases}\label{choiceofUsecond}
 \end{equation}
Notice that in both cases the proposed choice satisfies the Markov chain $\bar{U}_{\set{S},1} - \bar{U}_{2,\set{S}} - \cdots- \bar{U}_{\set{S},|\set{S}|-1} - X$.

Trivially, for $k\notin\big\{\tilde{k},\tilde{k}+1\big\}$, constraint \eqref{eq:d_inequa} is unchanged if we replace 
$(U_{\set{S},1}, U_{\set{S},2}, \ldots, U_{\set{S},|\set{S}|-1}, X)$ by  $(\bar{U}_{\set{S},1}, \bar{U}_{\set{S},2}, \ldots, \bar{U}_{\set{S},K-1}, {X})$ and $({\alpha}_{\set{S},1}, \ldots,{\alpha}_{\set{S},|\set{S}|})$ by  $(\bar{\alpha}_{\set{S},1}, \ldots, \bar{\alpha}_{\set{S},|\set{S}|})$. 

If \eqref{eq:condition1} holds, then the proposed replacement relaxes constraint~\eqref{eq:d_inequa}  for $k=\tilde{k}$ \big(because $\bar{\alpha}_{\set{S},\tilde{k}} >{\alpha}_{\set{S},\tilde{k}}$\big) and it tightens it  for $k=\tilde{k}+1$ \big(because $\bar{\alpha}_{\set{S},\tilde{k}+1} <{\alpha}_{\set{S},\tilde{k}+1}$\big). However, the new constraint for $k=\tilde{k}+1$ is less stringent than the original constraint for $k=\tilde{k}$:
\begin{IEEEeqnarray}{rCl}
\lefteqn{\bar{\alpha}_{\set{S},\tilde{k}+1}  + I(\bar{U}_{\set{S},\tilde{k}+1}; Y_{\tilde{k}+1}|\bar{U}_{\set{S},\tilde{k}}) } \quad \nonumber \\
&\stackrel{(a)}{=}& (1-\gamma)\cdot{\alpha}_{\set{S},\tilde{k}}+ \gamma\cdot {\alpha}_{\set{S},\tilde{k}+1}  +  I({U}_{\set{S},\tilde{k}+1}; Y_{\tilde{k}+1}|U_{\set{S},\tilde{k}}) \nonumber \\
& \stackrel{(b)}{>} &  {\alpha}_{\set{S},\tilde{k}} + I({U}_{\set{S},\tilde{k}+1}; Y_{\tilde{k}+1}|U_{\set{S},\tilde{k}}) \nonumber \\
&\stackrel{(c)}{\geq}  &  {\alpha}_{\set{S},\tilde{k}}   + I({U}_{\set{S},\tilde{k}}; Y_{\tilde{k}}|U_{\set{S},\tilde{k}-1}),
\end{IEEEeqnarray}
where (a) holds by \eqref{eq:barkplus}; (b) holds by \eqref{eq:strict}; and  (c)   holds by   holds by assumption~\eqref{eq:condition1}.
We conclude that when~\eqref{eq:condition1} holds,  the upper bound on $\C(\M_1, \ldots, \M_K)$ in \eqref{eq:d_inequa} is relaxed if everywhere one replaces \\$(U_{\set{S},1}, U_{\set{S},2} \ldots, U_{|\set{S},|\set{S}|-1})$ and $({\alpha}_{\set{S},1}, \ldots,{\alpha}_{\set{S},|\set{S}|})$  by $(\bar{U}_{\set{S},1}, \bar{U}_{\set{S},2}, \ldots, \bar{U}_{\set{S},|\set{S}|-1})$ and $(\bar{\alpha}_{\set{S},1}, \ldots, \bar{\alpha}_{\set{S},|\set{S}|})$.

We now assume that  \eqref{eq:condition2} holds. 
 We show that the new constraints obtained for $k=\tilde{k}$ and for $k=\tilde{k}+1$ cannot be more stringent then the tighter of the two original constraints for $k=\tilde{k}$ and  $k=\tilde{k}+1$. 
 
  Consider $k=\tilde{k}$. By \eqref{eq:beta} and \eqref{choiceofUsecond} we have  \begin{IEEEeqnarray}{rCl}\label{eq:ubareq}
  	\lefteqn{I(\bar{U}_{\set{S},\tilde{k}};Y_{\tilde{k}}|\bar{U}_{\set{S},\tilde{k}-1})} \quad \nonumber\\
  	&=&{I({U}_{\set{S},\tilde{k}};Y_{\tilde{k}}|\bar{U}_{\set{S},\tilde{k}-1}E)} \quad \nonumber\\	
  	&= &(1-\beta)\cdot I(U_{\set{S},\tilde{k}};Y_{\tilde{k}}| U_{\set{S},\tilde{k}-1} ) \nonumber \\
  	&  = & \gamma\cdot I(U_{\set{S},\tilde{k}};Y_{\tilde{k}}| U_{\set{S},\tilde{k}-1}) \nonumber \\
  	&& + (1-\gamma) \cdot I(U_{\set{S},\tilde{k}+1};Y_{\tilde{k}+1}|U_{\set{S},\tilde{k}}).\nonumber \\ 
  \end{IEEEeqnarray}By~\eqref{eq:bark} and \eqref{eq:ubareq}:
  \begin{IEEEeqnarray}{rCl}
\lefteqn{
 \bar{\alpha}_{\set{S},\tilde{k}}+ I(\bar{U}_{\set{S},\tilde{k}};Y_{\tilde{k}}| \bar{U}_{\set{S},\tilde{k}-1}) }\quad \nonumber \\
& = &\big( \gamma {\alpha}_{\set{S},\tilde{k}} + (1-\gamma)  {\alpha}_{\set{S},\tilde{k}+1} \big)  \nonumber \\
&&+ \gamma I({U}_{\set{S},\tilde{k}}; Y_{\tilde{k}}|U_{\set{S},\tilde{k}-1})+ (1-\gamma)  I({U}_{\set{S},\tilde{k}+1}; Y_{\tilde{k}+1}|U_{\set{S},\tilde{k}}) \nonumber \\
& \geq  & \min\big\{  {\alpha}_{\set{S},\tilde{k}}   + I({U}_{\set{S},\tilde{k}}; Y_{\tilde{k}}|U_{\set{S},\tilde{k}-1}), \nonumber \\
& & \hspace{2cm} {\alpha}_{\set{S},\tilde{k}+1}   + I({U}_{\set{S},\tilde{k}+1}; Y_{\tilde{k}+1}|U_{\set{S},\tilde{k}}) \big\}.
\end{IEEEeqnarray}

   Let now $k=\tilde{k}+1$. We have:
\begin{IEEEeqnarray}{rCl}
\lefteqn{
I(\bar{U}_{\set{S},\tilde{k}+1};Y_{\tilde{k}+1}| \bar{U}_{\set{S},\tilde{k}}) }\quad \nonumber \\
& \stackrel{(a)}{=} & (1-\beta) I(U_{\set{S},\tilde{k}+1};Y_{\tilde{k}+1}|{U}_{\set{S},\tilde{k}}) \nonumber \\
& & + \beta I(U_{\set{S},\tilde{k}+1};Y_{\tilde{k}+1}|{U}_{\set{S},\tilde{k}-1}) \nonumber \\
& \stackrel{(b)}{=} &  (1-\beta) I(U_{\set{S},\tilde{k}+1};Y_{\tilde{k}+1}|{U}_{\set{S},\tilde{k}}) \nonumber \\
& &+ \beta I(U_{\set{S},\tilde{k}+1}, U_{\set{S},\tilde{k}};Y_{\tilde{k}+1}|{U}_{\set{S},\tilde{k}-1}) \nonumber \\
& \stackrel{(c)}{=}  &  I(U_{\set{S},\tilde{k}+1};Y_{\tilde{k}+1}|{U}_{\set{S},\tilde{k}})+ \beta I( U_{\set{S},\tilde{k}};Y_{\tilde{k}+1}|{U}_{\set{S},\tilde{k}-1}) \nonumber \\
& \stackrel{(d)}{\geq}  &  I(U_{\set{S},\tilde{k}+1};Y_{\tilde{k}+1}|{U}_{\set{S},\tilde{k}})+ \beta I( U_{\set{S},\tilde{k}};Y_{\tilde{k}}|{U}_{\set{S},\tilde{k}-1}) \nonumber \\
& \stackrel{(e)}{=}  &\gamma I(U_{\set{S},\tilde{k}+1};Y_{\tilde{k}+1}|{U}_{\set{S},\tilde{k}})+ (1-\gamma) I( U_{\set{S},\tilde{k}};Y_{\tilde{k}}|{U}_{\set{S},\tilde{k}-1}),\nonumber \\
\end{IEEEeqnarray}
where (a) follows by the definition of $\bar{U}_{\set{S},\tilde{k}}$ and  $\bar{U}_{\set{S},\tilde{k}+1}$; (b) by the Markov chain \eqref{eq:markovsetS}; (c) by the chain rule of mutual information and  Markov chain \eqref{eq:markovsetS}; (d) by the degradedness of the channel \eqref{eq:markov_channel}; (e) by the definition of $\beta$ in \eqref{eq:beta}.

Therefore, by \eqref{eq:barkplus}:
\begin{IEEEeqnarray}{rCl}
\lefteqn{ \bar{\alpha}_{\set{S},k+1}+
I(\bar{U}_{\set{S},\tilde{k}+1};Y_{\tilde{k}+1}| \bar{U}_{\set{S},\tilde{k}}) } \nonumber \\
& \geq   &(1-\gamma) \cdot {\alpha}_{\set{S},\tilde{k}} + \gamma  \cdot{\alpha}_{\set{S},\tilde{k}+1}   \nonumber \\
&&+(1- \gamma )\cdot I({U}_{\set{S},\tilde{k}}; Y_{\tilde{k}}|U_{\set{S},\tilde{k}-1})+ \gamma \cdot I({U}_{\set{S},\tilde{k}+1}; Y_{\tilde{k}+1}|U_{\set{S},\tilde{k}}) \nonumber \\
& \geq  & \min\big\{  {\alpha}_{\set{S},\tilde{k}}   + I({U}_{\set{S},\tilde{k}}; Y_{\tilde{k}}|U_{\set{S},\tilde{k}-1}), \nonumber \\
& & \hspace{2cm} {\alpha}_{\set{S},\tilde{k}+1}   + I({U}_{\set{S},\tilde{k}+1}; Y_{\tilde{k}+1}|U_{\set{S},\tilde{k}}) \big\}.
\end{IEEEeqnarray}
We  thus conclude that also when \eqref{eq:condition2} holds,  the upper bound on $\C(\M_1, \ldots, \M_K)$ in \eqref{eq:d_inequa} is relaxed if one replaces $(U_{\set{S},1}, U_{\set{S},2}, \ldots, U_{\set{S},|\set{S}|-1})$ and $({\alpha}_{\set{S},1}, \ldots,{\alpha}_{\set{S},K})$  by $(\bar{U}_{\set{S},1}, \bar{U}_{\set{S}, 2}, \ldots, \bar{U}_{\set{S},|\set{S}|-1})$ and $(\bar{\alpha}_{\set{S},1}, \ldots, \bar{\alpha}_{\set{S},|\set{S}|})$.

\section{Proof of Remark~\ref{rem:1}} \label{app:proofremark}

We first prove that the bound  in Theorem~\ref{thm:upper_new} is loosened when each $\alpha_{\set{S},k}^\star$ is replaced by $\tilde{\alpha}_{\set{S},k}$.
Consider the intermediate Lemma~\ref{lem:intermediate} in the proof of Theorem~\ref{thm:upper_new}, Appendix~\ref{sec:upperbound}. 
Relax the upper bound in this lemma by replacing for $k=2,\ldots, K$ constraint~\eqref{eq:cons1} by 
\begin{equation}
\alpha_{\set{S},k} \geq0. 
\end{equation}
Following similar steps as in the proof of Lemma~\ref{lem:sym_alphas}, see also  \cite[Lemma~12]{saeeditimowigger-IT}, it can be shown that this relaxed upper bound is not changed when one imposes that
\begin{subequations}
\begin{IEEEeqnarray*}{Cl}
\alpha_{\set{S},2}= \alpha_{\set{S}, 3}=  \ldots=\alpha_{\set{S},|\set{S}|},
\end{IEEEeqnarray*}
and 
\begin{IEEEeqnarray*}{Cl}
\alpha_{\set{S},1}= \frac{\M_1}{D} \qquad \textnormal{or} \qquad 
\alpha_{\set{S},1}=\alpha_{\set{S},2}.
\end{IEEEeqnarray*}
\end{subequations}
Since constraints~\eqref{eq:d_inequa5} are increasing in $\alpha_{\set{S},1},\ldots, \alpha_{\set{S},|\set{S}|}$, by constraint~\eqref{eq:cons3prime}, we conclude  that the relaxed upper bound is loosest for 
\begin{IEEEeqnarray*}{rCl}
\alpha_{\set{S},1}&= &\frac{\M_1}{D}\\
\alpha_{\set{S},k} &= &\frac{|\set{S}| \sum_{i=1}^{|\set{S}|}\M_{j_i}-\M_{j_1}}{(|\set{S}|-1)D},  \quad k\in\{2,\ldots, |\set{S}|-1\},\IEEEeqnarraynumspace
\end{IEEEeqnarray*}
i.e., for $\alpha_{\set{S},k} = \tilde{\alpha}_{\set{S},k}$.

We now prove that the bound in Theorem~\ref{thm:upper_new} is loosened when each $\alpha_{\set{S},k}^\star$ is replaced by ${\alpha}_{\set{S},k}'$.
Consider again the intermediate Lemma~\ref{lem:intermediate} in Appendix~\ref{sec:upperbound}. Relax constraint \eqref{eq:cons1}  by replacing it with $\alpha_{\set{S},k} \geq 0$, for \emph{all} $k=1,\ldots, K$. Following the steps in \cite[Lemma~12]{saeeditimowigger-IT}, it can be shown that the new constraints are loosest if each 
\begin{equation}
\alpha_{\set{S},k}=\alpha_{\set{S},k}'.
\end{equation} This concludes the proof.
\section{Proof of Proposition \ref{cor:generalizedtight}}
\label{ap:cor:generalizedtight}

For $\Delta=0$,  achievability follows  by specializing Theorem~\ref{thm:multipg} to $t=K-1$ and to the input distribution $P_X$ that maximizes~\eqref{eq:max}. In fact, for this input distribution:
		\[R^{(K-1)}=  K\C_{\textnormal{avg}}=\C_{\textnormal{avg}}+ \frac{\sum_{k=1}^K \M_k^{\star(K-1)}}{ K \cdot D} .\]
		For $\Delta>0$, achievability follows from Proposition~\ref{prop:local}.
		
The converse is proved as follows. Apply Theorem~\ref{thm:upper_new}, but consider only the  constraints~\eqref{eq:d_inequa10} corresponding to the sets $\set{S}=\{k\}$, for $k\in \K$. Taking the average over the resulting $K$ constraints, establishes that there exists a random variable $(X, Y_1, \ldots, Y_K)$ satisfying \eqref{eq:markov_channel} and so that 
\begin{IEEEeqnarray}{rCl}
\C(\M_1, \ldots, \M_K) \leq \frac{1}{K} \sum_{k\in\K} I(X;Y_k) + \frac{1}{K} \sum_{k\in\K} \frac{\M_k}{D}.\IEEEeqnarraynumspace
\end{IEEEeqnarray}
Maximizing the right-hand side over input distributions $P_X$ yields the desired converse.

\section{Proof of Proposition \ref{prop:loosened}}
\label{ap:smallMnotpossible}

Relax the upper bound in Theorem~\ref{thm:upper_new} by   considering  constraints \eqref{eq:d_inequa10} only for the set of all receivers $\set{S}=\K$, 
and by replacing each $\alpha_{\set{S},k}^\star$ by $\tilde{\alpha}_{\set{S},k}$.
Specializing the resulting relaxed bound to the erasure BC, one obtains
 the  following upper bound: 
\begin{IEEEeqnarray}{rCl}\label{eq:Cu}
\C(\M_1, \ldots, \M_K)& \leq&  \max \min\bigg\{ (1-\delta_1) \beta_1 + \frac{\M_1}{D},  (1-\delta_2) \beta_2 + \frac{K \M- \M_1}{D \cdot (K-1)}, \ldots, (1-\delta_K) \beta_K +\frac{K \M- \M_1}{D \cdot (K-1)}\bigg\}, \nonumber \\
\end{IEEEeqnarray}
where the maximization is over the choice of parameters $\beta_1 ,  \beta_2 , \ldots , \beta_K\geq 0$ satisfying
\begin{equation}\label{eq:betas}
\sum_{k=1}^K \beta_k \leq 1. 
\end{equation}

The upper bound in the proposition is established by solving this maximization  problem. In fact, by noticing that the bound is increasing in $\beta_1 ,  \beta_2 , \ldots , \beta_K\geq 0$, and by first fixing $\beta_1$ and optimizing over the choices $\beta_2, \ldots, \beta_K \geq 0 $ summing to $1-\beta_1$, we obtain
\begin{IEEEeqnarray}{rCl}
	\lefteqn{
\C(\M_1, \ldots, \M_K)} \; \nonumber \\
&  \leq&  \max_{\beta_1\in[0,1]} \min\bigg\{ \beta_1\C_1  + \frac{\M_1}{D}, \nonumber \\
& & \qquad \qquad \qquad  (1-\beta_1)\C_{\{2,\ldots, K\}}  +   \frac{K \M- \M_1}{(K-1)\cdot D}\bigg\}, \nonumber\\
& =&  \max_{\beta_1\in[0,1]} \min\bigg\{ \beta_1\C_1  , 
  (1-\beta_1)\C_{\{2,\ldots, K\}}  +   \frac{K( \M- \M_1)}{ (K-1)\cdot D}\bigg\} \nonumber \\&& \quad + \frac{\M_1}{D}.  \nonumber\\
\end{IEEEeqnarray}
If 
\[ \frac{K( \M- \M_1)}{ (K-1)\cdot D} \geq \C_1,
\] then the maximum is achieved at $\beta_1=1$ and the upper bound results in 
\begin{equation}
\C(\M_1, \ldots, \M_K) \leq \C_1 + \frac{\M_1}{D}. 
\end{equation}
Otherwise the maximum is at $\beta=\beta^\star$, where
\begin{equation}
\beta_1^\star:= \frac{\C_{\{2,\ldots, K\}}+\frac{K( \M- \M_1)}{ (K-1)\cdot D} }{\C_1 + \C_{\{2,\ldots, K\}}},
\end{equation}
and the upper bound results in 
\begin{IEEEeqnarray}{rCl}
\C(\M_1, \ldots, \M_K) &\leq& \C_{\K} + \frac{K( \M- \M_1)}{ (K-1)\cdot D} \cdot \frac{\C_1}{\C_1+ \C_{\{2,\ldots, K\}}} +\frac{\M_1}{D},\nonumber\\
& = & \C_{\K} + \frac{K( \M- \M_1)}{ (K-1)\cdot D} \cdot \frac{\C_{\K}}{ \C_{\{2,\ldots, K\}}} +\frac{\M_1}{D},
\end{IEEEeqnarray}
where we used that for erasure BCs
\begin{equation}
\C_\K= \frac{ \C_1 \cdot \C_{\{2,\ldots, K\}}}{\C_1 +  \C_{\{2,\ldots, K\}}}.
	\end{equation}

\bigskip

\medskip

\section{Proof of Corollary~\ref{cor:minimumrate}} \label{app:corBEC}
Fix $t \in \K$ and $\set{S}=\{1,\ldots, t\}$. 
For the considered channel
\begin{equation}
(r_1 \ldots, r_t) \in \capa_{\set{S}} \quad \Longleftrightarrow \quad \sum_{k=1}^t r_k\leq 1.
\end{equation}
		The upper bound in Corollary~\ref{cor-equivalent} thus states that for this noise-free BC a rate-memory tuple $(R, \M_1, \ldots, \M_K)$ is achievable only if 
		\begin{equation}
t R - \sum_{k=1}^t \alpha_{\set{S},k}^\star  \leq 1.
		\end{equation}
		This is equivalent to the following bound on the capacity-memory tradeoff 
		\begin{equation}\label{eq:Cup}
		\C(\M_1, \ldots,\M_K) \leq \frac{1}{t} \left( 1+ \sum_{k=1}^t \alpha_{\set{S},k}^\star \right). 
		\end{equation}
		Notice that the sum  $\sum_{k=1}^t \alpha_{\set{S},k}^\star$ takes on only two different values, depending on the outcomes of the minimizations defining $\alpha_{\set{S},k}^{\star}$. It is either
		\begin{subequations}\label{eq:sumalpha}
		\begin{equation} 
		\sum_{k=1}^t \alpha_{\set{S},k}^\star=\frac{t \sum_{k=1}^t \M_{k}}{D}
		\end{equation}
		or 
			\begin{equation} 
		\sum_{k=1}^t \alpha_{\set{S},k}^\star=
		\sum_{k=1}^{t} \frac{\sum_{i=1}^{k} \M_{i}}{D- k+1}.
		\end{equation}
		\end{subequations}
		
		Combining \eqref{eq:Cup} with \eqref{eq:sumalpha}, applying the correspondence $\rho=R^{-1}$ and $m_k=\frac{\M_k}{R}$, and setting $m_1=m_2=\ldots=m_k=m$ yields, 
		\begin{equation}
		1 \leq \frac{1}{t} \left( \rho + m\cdot\min\bigg\{\frac{t^2 }{D},\;\sum_{k=1}^{t} \frac{k}{D-k+1} \bigg\}\right),
		\end{equation}
		which is equivalent to the bound in the corollary. 

\end{document}

%% file: ErasureBC.tex
\centering
\begin{tikzpicture} [every pin/.style={fill=white},scale=0.9]
  \begin{axis}[scale=1.3,
width=0.5\textwidth,
scale only axis,
xmin=0,
xmax=8,
xmajorgrids,
xlabel={\Large{ $\frac{\M}{D}$}},
ymin=0,
ymax=2.75,
ymajorgrids,
ylabel={\large{Rate}},
axis x line*=bottom,
axis y line*=left,
legend pos=north west,
legend style={draw=none,fill=none,legend cell align=left, font=\large}
]

         \addplot[color=magenta,dashed,line width=1pt]
 table[row sep=crcr]{  
                   0   0.068195413758724\\
   0.100000000000000   0.109693877551020\\
   0.200000000000000   0.142222222222222\\
   0.300000000000000   0.173333333333333\\
   0.400000000000000   0.200000000000000\\
   0.500000000000000   0.225000000000000\\
   0.600000000000000   0.250000000000000\\
   0.700000000000000   0.275000000000000\\
   0.800000000000000   0.300000000000000\\
   0.900000000000000   0.325000000000000\\
   1.000000000000000   0.350000000000000\\
   1.100000000000000   0.375000000000000\\
   1.200000000000000   0.400000000000000\\
   1.300000000000000   0.425000000000000\\
   1.400000000000000   0.450000000000000\\
   1.500000000000000   0.475000000000000\\
   1.600000000000000   0.500000000000000\\
   1.700000000000000   0.525000000000000\\
   1.800000000000000   0.550000000000000\\
   1.900000000000000   0.575000000000000\\
   2.000000000000000   0.600000000000000\\
   2.100000000000000   0.625000000000000\\
   2.200000000000000   0.650000000000000\\
   2.300000000000000   0.675000000000000\\
   2.400000000000000   0.700000000000000\\
   2.500000000000000   0.725000000000000\\
   2.600000000000000   0.750000000000000\\
   2.700000000000000   0.775000000000000\\
   2.800000000000000   0.800000000000000\\
   2.900000000000000   0.825000000000000\\
   3.000000000000000   0.850000000000000\\
   3.100000000000000   0.875000000000000\\
   3.200000000000000   0.900000000000000\\
   3.300000000000000   0.925000000000000\\
   3.400000000000000   0.950000000000000\\
   3.500000000000000   0.975000000000000\\
   3.600000000000000   1.000000000000000\\
   3.700000000000000   1.025000000000000\\
   3.800000000000000   1.050000000000000\\
   3.900000000000000   1.075000000000000\\
   4.000000000000000   1.100000000000000\\
   4.100000000000000   1.125000000000000\\
   4.200000000000000   1.150000000000000\\
   4.300000000000000   1.175000000000000\\
   4.400000000000000   1.200000000000000\\
   4.500000000000000   1.225000000000000\\
   4.600000000000000   1.250000000000000\\
   4.700000000000000   1.275000000000000\\
   4.800000000000000   1.300000000000000\\
   4.900000000000000   1.325000000000000\\
   5.000000000000000   1.350000000000000\\
   5.100000000000000   1.375000000000000\\
   5.200000000000000   1.400000000000000\\
   5.300000000000000   1.425000000000000\\
   5.400000000000000   1.450000000000000\\
   5.500000000000000   1.475000000000000\\
   5.600000000000000   1.500000000000000\\
   5.700000000000000   1.525000000000000\\
   5.800000000000000   1.550000000000000\\
   5.900000000000000   1.575000000000000\\
   6.000000000000000   1.600000000000000\\
   6.100000000000000   1.625000000000000\\
   6.200000000000000   1.650000000000000\\
   6.300000000000000   1.675000000000000\\
   6.400000000000000   1.700000000000000\\
   6.500000000000000   1.725000000000000\\
   6.600000000000000   1.750000000000000\\
   6.700000000000000   1.775000000000000\\
   6.800000000000000   1.800000000000000\\
   6.900000000000000   1.825000000000000\\
   7.000000000000000   1.850000000000000\\};
    \addlegendentry{Upper bound under uniform cache-assignment}

 \addplot[color=red,solid,line width=1pt]
 table[row sep=crcr]{
                        0   0.068195413758472\\
   0.100000000000000   0.168195413562149\\
   0.200000000000000   0.257210031343816\\
   0.300000000000000   0.307210027932690\\
   0.400000000000000   0.357210030989821\\
   0.500000000000000   0.407210031334816\\
   0.600000000000000   0.457210031342373\\
   0.700000000000000   0.507210031343710\\
   0.800000000000000   0.557210030891561\\
   0.900000000000000   0.604113344113301\\
   1.000000000000000   0.644113342846768\\
   1.100000000000000   0.684113344113342\\
   1.200000000000000   0.720720720720718\\
   1.300000000000000   0.754054054054052\\
   1.400000000000000   0.787387387387385\\
   1.500000000000000   0.820720720720722\\
   1.600000000000000   0.854054054054053\\
   1.700000000000000   0.887387384794337\\
   1.800000000000000   0.920720719198443\\
   1.900000000000000   0.954054053116297\\
   2.000000000000000   0.987387386809711\\
   2.100000000000000   1.020720720364460\\
   2.200000000000000   1.054054053823501\\
   2.300000000000000   1.087387387357211\\
   2.400000000000000   1.120720720700230\\
   2.500000000000000   1.154054054038562\\
   2.600000000000000   1.187387387373927\\
   2.700000000000000   1.220720720706250\\
   2.800000000000000   1.254054054031881\\
   2.900000000000000   1.287387387348732\\
   3.000000000000000   1.320720720289332\\
   3.100000000000000   1.354054052900109\\
   3.200000000000000   1.387387387368721\\
   3.300000000000000   1.412499999999573\\
   3.400000000000000   1.437499999999921\\
   3.500000000000000   1.462499999440169\\
   3.600000000000000   1.487499999976217\\
   3.700000000000000   1.512499999986614\\
   3.800000000000000   1.537499999992137\\
   3.900000000000000   1.562499999999510\\
   4.000000000000000   1.587499999999673\\
   4.100000000000000   1.612499999999774\\
   4.200000000000000   1.637499999999832\\
   4.300000000000000   1.662499999999873\\
   4.400000000000000   1.687499999999899\\
   4.500000000000000   1.712499999999915\\
   4.600000000000000   1.737499999999929\\
   4.700000000000000   1.762499999999942\\
   4.800000000000000   1.787499999999998\\
   4.900000000000000   1.812499999999994\\
   5.000000000000000   1.837499999999998\\
   5.100000000000000   1.862499999999995\\
   5.200000000000000   1.887499999999998\\
   5.300000000000000   1.912499999999997\\
   5.400000000000000   1.937499999999998\\
   5.500000000000000   1.962499999999999\\
   5.600000000000000   1.987499999999995\\
   5.700000000000000   2.012499999999998\\
   5.800000000000000   2.037500000000000\\
   5.900000000000000   2.062500000000003\\
   6.000000000000000   2.087500000000001\\
   6.100000000000000   2.112499993561400\\
   6.200000000000000   2.137499994573904\\
   6.300000000000000   2.162499995317630\\
   6.400000000000000   2.187499995765894\\
   6.500000000000000   2.212499996230696\\
   6.600000000000000   2.237499996687624\\
   6.700000000000000   2.262499997121288\\
   6.800000000000000   2.287499997496794\\
   6.900000000000000   2.312499997764312\\
   7.000000000000000   2.337499998012994\\
   7.100000000000000   2.362499998221829\\
   7.200000000000000   2.387499998400251\\
   7.300000000000000   2.412499998402034\\
   7.400000000000000   2.437499998370872\\
   7.500000000000000   2.462499998366324\\
   7.600000000000000   2.487499998384004\\
   7.700000000000000   2.512499998416297\\
   7.800000000000000   2.537499998459409\\
   7.900000000000000   2.562499998510329\\
   8.000000000000000   2.587499998566626\\};
    \addlegendentry{Upper bound on $\C^\star(\M)$}

 \addplot[color=blue, solid,line width=1pt,mark=*]
 table[row sep=crcr]{
   0   0.068195413758724\\
   0.068195413758724 0.136390827517448\\
   0.275549450549451   0.275549450549451\\
   1.548936170212766   0.774468085106383\\
   7.050000000000001   2.350000000000000\\};
    \addlegendentry{Lower bound on  $\C^\star(\M)$}

\end{axis}

\end{tikzpicture}

%% file: K12D64_sourcecoding.tex
\begin{figure}[h!]
\centering
\begin{tikzpicture} [every pin/.style={fill=white},scale=0.9]
  \begin{axis}[scale=1.3,
width=0.5\textwidth,
scale only axis,
xmin=0,
xmax=64,
xmajorgrids,
xlabel={\large{Memory $\M_1=\ldots=\M_K$}},
ymin=0,
ymax=12,
ymajorgrids,
ylabel={\large{Minimum delivery rate $\rho$}},
axis x line*=bottom,
axis y line*=left,
legend pos=north east,
legend style={draw=none,fill=none,legend cell align=left, font=\large}
]

         \addplot[color=blue,solid,line width=1]
 table[row sep=crcr]{  
         0    12.00000 \\
    1    10.61995\\
    2    9.23990\\
    3    7.85984\\
    4    6.47979\\
    5    5.25033\\
    6    4.39131\\
    7    3.74820\\
    8    3.26360\\
    9    2.88276\\
    10    2.55335\\
     11    2.30868\\
    12    2.06402\\
    13    1.90269\\
    14    1.74135\\
    15    1.58002\\
    16    1.46787\\
    17    1.37211\\
    18    1.27635\\
    19    1.18060\\
    20    1.08484\\
        21    1.00521\\
    22    0.95784\\
    23    0.91047\\
    24    0.86310\\
    25    0.81572\\
    26    0.76835\\
    27    0.72098\\
    28    0.67361\\
    29    0.62624\\
    30    0.57887\\
        31    0.53150\\
    32    0.50000\\
    33    0.48438\\
    34    0.46875 \\
    35    0.45312\\
    36    0.43750\\
    37    0.42188\\
    38    0.40625\\
    39    0.39062\\
    40    0.37500\\
        41    0.35938\\
    42    0.34375\\
    43    0.32812\\
    44    0.31250\\
    45    0.29688\\
    46    0.28125\\
    47    0.26562\\
    48    0.25000\\
    49    0.23438\\
    50    0.21875\\
    51    0.20312                                       \\
    52    0.18750\\
    53    0.17188\\
    54    0.15625\\
    55    0.1267\\
    56    0.12500\\
    57    0.10938\\
    58    0.09375\\
    59    0.07812\\
    60    0.06250\\
    61    0.04688\\
    62    0.03125\\
    63    0.01562\\
        64    0\\
    };
    \addlegendentry{Lower bound of Cor.~\ref{cor:minimumrate}}

  \addplot[color=magenta,only marks,width=1pt, mark=*]
 table[row sep=crcr]{
      5.33333    4.3611\\};
    \addlegendentry{Lower bound of \cite{ghasemi_ramamoorthy}}

 \addplot[color=black,dashed,line width=1pt]
 table[row sep=crcr]{
   0    11.02062                     \\
    1    8.95425\\
    2    6.88789\\
    3    4.95411\\
    4    3.78795\\
    5    3.06555\\
    6    2.57453\\
    7    2.19781\\
    8    1.95361\\
    9    1.70941\\
    10    1.56898\\
     11    1.43054\\
    12    1.29210\\
    13    1.17822\\
    14    1.11621\\
    15    1.05420\\
    16    0.99219\\
    17     0.93018                                          
    18    0.86816\\
    19    0.80615\\
    20    0.74414\\
        21    0.68213\\
    22    0.65625\\
    23    0.64062\\
    24    0.62500\\
    25    0.60938\\
    26    0.59375\\
    27    0.57812\\
    28    0.56250\\
    29    0.54688\\
    30    0.53125\\
        31    0.51562\\
    32    0.50000\\
    33    0.48438\\
    34    0.46875 \\
    35    0.45312\\
    36    0.43750\\
    37    0.42188\\
    38    0.40625\\
    39    0.39062\\
    40    0.37500\\
        41    0.35938\\
    42    0.34375\\
    43    0.32812\\
    44    0.31250\\
    45    0.29688\\
    46    0.28125\\
    47    0.26562\\
    48    0.25000\\
    49    0.23438\\
    50    0.21875\\
    51    0.20312                                       \\
    52    0.18750\\
    53    0.17188\\
    54    0.15625\\
    55    0.1267\\
    56    0.12500\\
    57    0.10938\\
    58    0.09375\\
    59    0.07812\\
    60    0.06250\\
    61    0.04688\\
    62    0.03125\\
    63    0.01562\\
        64    0\\
    };
    \addlegendentry{Lower bound of \cite{wanglimgastpar-2016}}

 \addplot[color=red,solid,line width=0.7pt,mark=+]
 table[row sep=crcr]{
 0 12.00000   \\
 5.33333  5.50000   \\
  10.66666   3.33333  \\
   16  2.25000   \\ 
  21.33333   1.60000   \\ 
   26.66666  1.16667   \\
  32    0.85714   \\
   37.33333    0.62500   \\
   42.66666      0.44444 \\
     48      0.30000  \\
     53.33333        0.18182 \\
      58.66666           0.08333 \\
        64           0.00000
  \\};
    \addlegendentry{Upper bound of \cite{yumaddahaliavestimehr-2016}}

\end{axis}
\end{tikzpicture}

\caption{Upper and lower bounds on the minimum delivery  rate $\rho$ in the source coding model of \cite{maddahali_niesen_2014-1} for $K=12$ and $D=64$.}
\label{fig:source}

\end{figure}

%% file: K4D10Gaussian.tex
\centering
\begin{tikzpicture} [every pin/.style={fill=white},scale=0.9]
  \begin{axis}[scale=1.2,
width=0.5\textwidth,
scale only axis,
xmin=0,
xmax=10,
xmajorgrids,
xlabel={\Large{$\frac{\M}{D}$}},
ymin=0,
ymax=3.5,
ymajorgrids,
ylabel={\large{Rate}},
axis x line*=bottom,
axis y line*=left,
legend pos=north west,
legend style={draw=none,fill=none,legend cell align=left, font=\large}
]

         \addplot[color=magenta,dashed,line width=1pt]
 table[row sep=crcr]{  
     0   0.095275309732723\\
   0.050000000000000   0.116462846277446\\
   0.100000000000000   0.136744169566951\\
   0.150000000000000   0.157714659462889\\
   0.200000000000000   0.176776968981119\\
   0.250000000000000   0.194833743774855\\
   0.300000000000000   0.212616571589958\\
   0.350000000000000   0.228365859860666\\
   0.400000000000000   0.246053184378514\\
   0.450000000000000   0.263755421032985\\
   0.500000000000000   0.282846234102109\\
   0.550000000000000   0.298463987443685\\
   0.600000000000000   0.310963987443688\\
   0.650000000000000   0.323463987443685\\
   0.700000000000000   0.335963987443685\\
   0.750000000000000   0.348463987443684\\
   0.800000000000000   0.360963987443685\\
   0.850000000000000   0.373463987443686\\
   0.900000000000000   0.385963987443688\\
   0.950000000000000   0.398463987443688\\
   1.000000000000000   0.410963987443686\\
   1.050000000000000   0.423463987443688\\
   1.100000000000000   0.435963987443686\\
   1.150000000000000   0.448463987443686\\
   1.200000000000000   0.460963987443688\\
   1.250000000000000   0.473463987443686\\
   1.300000000000000   0.485963987443688\\
   1.350000000000000   0.498463987443686\\
   1.400000000000000   0.510963987443688\\
   1.450000000000000   0.523463987443688\\
   1.500000000000000   0.535963987443685\\
   1.550000000000000   0.548463987443685\\
   1.600000000000000   0.560963987443688\\
   1.650000000000000   0.573463987443688\\
   1.700000000000000   0.585963987443688\\
   1.750000000000000   0.598463987443688\\
   1.800000000000000   0.610963987443688\\
   1.850000000000000   0.623463987443685\\
   1.900000000000000   0.635963987443685\\
   1.950000000000000   0.648463987443685\\
   2.000000000000000   0.660963987443685\\
   2.050000000000000   0.673463987443685\\
   2.100000000000000   0.685963987443685\\
   2.150000000000000   0.698463987443685\\
   2.200000000000000   0.710963987443685\\
   2.250000000000000   0.723463987443685\\
   2.300000000000000   0.735963987443685\\
   2.350000000000000   0.748463987443685\\
   2.400000000000000   0.760963987443685\\
   2.450000000000000   0.773463987443685\\
   2.500000000000000   0.785963987443685\\
   2.550000000000000   0.798463987443685\\
   2.600000000000000   0.810963987443685\\
   2.650000000000000   0.823463987443685\\
   2.700000000000000   0.835963987443685\\
   2.750000000000000   0.848463987443685\\
   2.800000000000000   0.860963987443685\\
   2.850000000000000   0.873463987443685\\
   2.900000000000000   0.885963987443685\\
   2.950000000000000   0.898463987443685\\
   3.000000000000000   0.910963987443685\\
   3.050000000000000   0.923463987443685\\
   3.100000000000000   0.935963987443685\\
   3.150000000000000   0.948463987443685\\
   3.200000000000000   0.960963987443685\\
   3.250000000000000   0.973463987443685\\
   3.300000000000000   0.985963987443685\\
   3.350000000000000   0.998463987443685\\
   3.400000000000000   1.010963987443685\\
   3.450000000000000   1.023463987443685\\
   3.500000000000000   1.035963987443685\\
   3.550000000000000   1.048463987443685\\
   3.600000000000000   1.060963987443685\\
   3.650000000000000   1.073463987443685\\
   3.700000000000000   1.085963987443685\\
   3.750000000000000   1.098463987443685\\
   3.800000000000000   1.110963987443685\\
   3.850000000000000   1.123463987443685\\
   3.900000000000000   1.135963987443685\\
   3.950000000000000   1.148463987443685\\
   4.000000000000000   1.160963987443685\\
   9.500000000000000   2.535963987443695\\};
    \addlegendentry{Upper bound under equal cache assignment}

 \addplot[color=red,solid,line width=1pt]
 table[row sep=crcr]{
                   0   0.111099406886968\\
   0.100000000000000   0.211099401296452\\
   0.200000000000000   0.311099401296452\\
   0.300000000000000   0.377983587552146\\
   0.400000000000000   0.427983587557030\\
   0.500000000000000   0.477983587555555\\
   0.600000000000000   0.527983587557295\\
   0.700000000000000   0.577983587557266\\
   0.800000000000000   0.627983587502271\\
   0.900000000000000   0.677983587557120\\
   1.000000000000000   0.727983585851775\\
   1.100000000000000   0.777983582585807\\
   1.200000000000000   0.820797700718771\\
   1.300000000000000   0.860006933180288\\
   1.400000000000000   0.893340268134912\\
   1.500000000000000   0.926673615621094\\
   1.600000000000000   0.960006550677340\\
   1.700000000000000   0.993340282990869\\
   1.800000000000000   1.026673591741726\\
   1.900000000000000   1.060006944617077\\
   2.000000000000000   1.093340284788137\\
   2.100000000000000   1.126673618101989\\
   2.200000000000000   1.160006886430079\\
   2.300000000000000   1.193340284532329\\
   2.400000000000000   1.226673618118473\\
   2.500000000000000   1.260006951247346\\
   2.600000000000000   1.293340284705343\\
   2.700000000000000   1.326673618017102\\
   2.800000000000000   1.360006951331314\\
   2.900000000000000   1.393340282852356\\
   3.000000000000000   1.426673618061579\\
   3.100000000000000   1.460006951402590\\
   3.200000000000000   1.493340251515006\\
   3.300000000000000   1.526673617337454\\
   3.400000000000000   1.560006951410019\\
   3.500000000000000   1.593340282725215\\
   3.600000000000000   1.626673617666431\\
   3.700000000000000   1.660006951236164\\
   3.800000000000000   1.693340188313010\\
   3.900000000000000   1.726673617508189\\
   4.000000000000000   1.760006951389598\\
   4.100000000000000   1.793340279229032\\
   4.200000000000000   1.826673616398619\\
   4.300000000000000   1.860006951256404\\
   4.400000000000000   1.893340283324811\\
   4.500000000000000   1.920790274627455\\
   4.600000000000000   1.945790240943789\\
   4.700000000000000   1.970789683253815\\
   4.800000000000000   1.995787716946939\\
   4.900000000000000   2.020790270360879\\
   5.000000000000000   2.045790257652869\\
   5.100000000000000   2.070790275844237\\
   5.200000000000000   2.095790271764778\\
   5.300000000000000   2.120790237992292\\
   5.400000000000000   2.145790244697451\\
   5.500000000000000   2.170790275216953\\
   5.600000000000000   2.195790275418776\\
   5.700000000000000   2.220790275167802\\
   5.800000000000000   2.245790270966279\\
   5.900000000000000   2.270790260949436\\
   6.000000000000000   2.295790274852178\\
   6.100000000000000   2.320790267606740\\
   6.200000000000000   2.345790270043249\\
   6.300000000000000   2.370790270537222\\
   6.400000000000000   2.395790267591535\\
   6.500000000000000   2.420790259221507\\
   6.600000000000000   2.445790244324574\\
   6.700000000000000   2.470790225426788\\
   6.800000000000000   2.495789921138801\\
   6.900000000000000   2.520790268561863\\
   7.000000000000000   2.545790268366901\\
   7.100000000000000   2.570790273334616\\
   7.200000000000000   2.595789358280083\\
   7.300000000000000   2.620790256994987\\
   7.400000000000000   2.645790236201411\\
   7.500000000000000   2.670790268937314\\
   7.600000000000000   2.695790271110053\\
   7.700000000000000   2.720790248151682\\
   7.800000000000000   2.745789839186326\\
   7.900000000000000   2.770790264295468\\
   8.000000000000000   2.795790156100919\\
   8.100000000000000   2.820790274383552\\
   8.199999999999999   2.845790254282050\\
   8.300000000000001   2.870790176796072\\
   8.400000000000000   2.895790273713009\\
   8.500000000000000   2.920790232026435\\
   8.600000000000000   2.945790275206517\\
   8.699999999999999   2.970790239628682\\
   8.800000000000001   2.995789199370863\\
   8.900000000000000   3.020790256818190\\
   9.000000000000000   3.045790272587531\\
   9.100000000000000   3.070790129817768\\
   9.199999999999999   3.095790251893882\\
   9.300000000000001   3.120790241503431\\
   9.400000000000000   3.145790059240918\\
   9.500000000000000   3.170790259533067\\
   9.600000000000000   3.195790234186842\\
   9.699999999999999   3.220790239709532\\
   9.800000000000001   3.245790267648006\\
   9.900000000000000   3.270790270484635\\
  10.000000000000000   3.295790270853648\\
  10.100000000000000   3.320790209937186\\
  10.199999999999999   3.345790275527677\\
  10.300000000000001   3.370790248925894\\
  10.400000000000000   3.395790242523975\\
  10.500000000000000   3.373513456292248\\};
    \addlegendentry{Upper bound on $\C^\star(\M)$}

 \addplot[color=blue, solid,line width=1pt,mark=*]
 table[row sep=crcr]{
             0    0.111099401296452\\
0.0690976372710104            0.180197038567462\\
          0.355645207662285   0.355645207662285\\
    1.959266806255855    0.979633403127928\\
    9.549483321368724    3.183161107122908\\};
    \addlegendentry{Lower bound on  $\C^\star(\M)$}

\end{axis}

\end{tikzpicture}